\documentclass[floatfix,reprint,aip,jcp,amsmath,citeautoscript]{revtex4-1}

\usepackage{graphicx}
\usepackage[hidelinks]{hyperref}
\usepackage{braket}
\usepackage{amsmath}	
\DeclareMathOperator{\trace}{tr}			

\usepackage{txfonts}

\begin{document}

\title{Overcoming positivity violations for density matrices in surface hopping} 

\author{A. S. Bondarenko}
\affiliation{$^{1)}$Department of Chemistry, Northwestern University, 2145 Sheridan Road, Evanston, Illinois 60208, USA}

\author{R. Tempelaar}
\email[]{roel.tempelaar@northwestern.edu}
\affiliation{$^{1)}$Department of Chemistry, Northwestern University, 2145 Sheridan Road, Evanston, Illinois 60208, USA}

\begin{abstract}
Fewest-switches surface hopping (FSSH) has emerged as one of the leading methods for modeling the quantum dynamics of molecular systems. While its original formulation was limited to adiabatic populations, the growing interest in the application of FSSH to coherent phenomena prompts the question how one should construct a complete density matrix based on FSSH trajectories. A straightforward solution is to define adiabatic coherences based on wavefunction coefficients. In this Paper, we demonstrate that inconsistencies introduced in the density matrix through such treatment may lead to a violation of positivity. We furthermore show that a recently proposed coherent generalization of FSSH results in density matrices that satisfy positivity, while yielding an improved accuracy throughout much (but not all) of parameter space.
\end{abstract}

\maketitle

\section{Introduction}

Many dynamical phenomena in chemistry and physics are effectively described by partitioning the problem of interest into a system and its environment, while limiting an explicit quantum treatment to the system in the form of a reduced density matrix. Given that appropriate approximations are made, this enables the accurate computational modeling of phenomena that are otherwise intractable. The multitude of approximations that can be taken has led to a wide variety of quantum dynamical methods that have been proposed over the years. When invoking approximations, in addition to the sheer accuracy of the method, it is of particular importance that the physical properties of the (reduced) density matrix are conserved. A well-known example of a violation of such physical properties is the breaking of positivity of the density matrix under the Markov approximation \cite{Dumcke.ZPhys.1979, Spohn.RevModPhys.1980, Pechukas.chapter.1991}. Both the Nakajima-Zwanzig and Redfield master equations violate positivity in principle \cite{Fay.JCP.2019}, with manifestations of this being particularly well documented for the latter \cite{Suarez.JCP.1992, Cheng.JCPB.2005, Ishizaki.adequacy_of_redfield.JCP.2009}. Within Redfield theory, positivity can be preserved by additionally invoking the secular approximation \cite{Redfield.IBM.1957, Redfield.AdvMagnoptRes.1965, May.Kuhn.book.2000}, albeit at the cost of additional approximations.

In 1990 Tully introduced fewest-switches surface hopping (FSSH) \cite{Tully.JCP.1990}, which has grown into one of the most widely-used quantum dynamical methods \cite{Nelson.ACR.2014, Subotnik.ARPC.2016, Wang.JPCL.2016}. FSSH is a mixed quantum--classical technique that represents the quantum system by means of a swarm of ``active'' surfaces that are allowed to switch between adiabats (instantaneous eigenstates of the quantum Hamiltonian) as a result of interactions with a classical environment. Intended for the modeling of (incoherent) scattering phenomena \cite{Tully.JCP.1990, Hammes.JCP.1994}, FSSH in its original form exclusively specified adiabatic populations. However, a growing interest in the application of FSSH to coherent phenomena \cite{Tempelaar.JCP.2013, Tempelaar.mapping.JPCL.2014, Petit.JCP.2014, Zimmermann.JCP.2014, vanderVegte.JPCB.2014, vanderVegte.JPCB.2015, Petit.JCTC.2015, Chen.JCP.2016, Richter.JCTC.2016, Hoffmann.JCP.2019, Jain.JPCB.2019, Sindhu.CPC.2022} prompted the question of how to construct the entire density matrix based on FSSH. Arguably the most straightforward and common approach to construct the coherent density matrix elements is by invoking the wavefunction coefficients that are propagated through the time-dependent Schr\"odinger equation, and which govern the hopping probabilities \cite{Tempelaar.JCP.2013, Landry.JCP.2013}. Although it was immediately recognized that this approach (henceforth simply referred to as FSSH) yields inconsistencies between populations and coherences, only recently a systematic survey of the spin-boson model has shown that it introduces inaccuracies in some cases, although no unphysical behavior was observed \cite{Tempelaar.CSH.JCP.2018}.

\begin{figure}[b]
\includegraphics{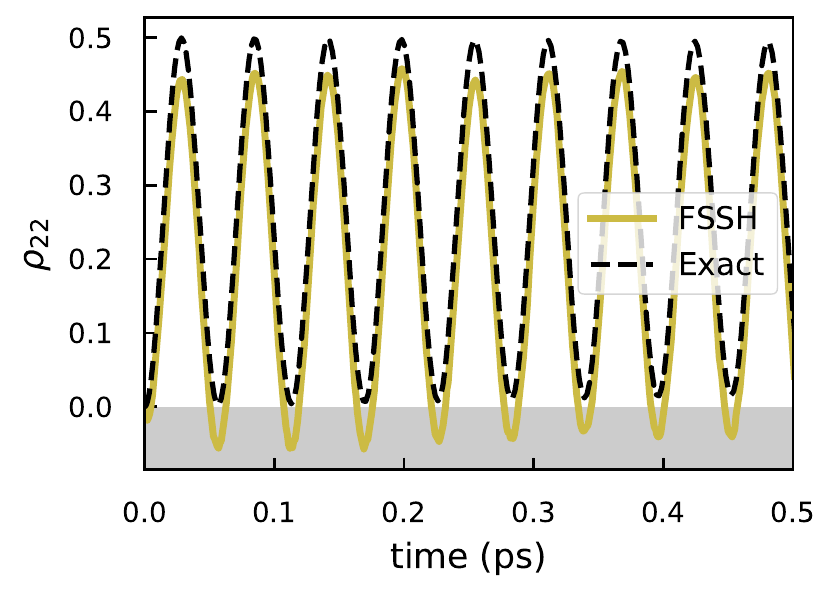}

\caption{Example of a violation of positivity by FSSH. Shown is the transient population of the second level of a homogeneous trimer in the local basis (see Sec.~\ref{sec:model} for details). Region of negative populations is highlighted in grey for ease of demonstration. Applied parameter are $V=1.0$, $\lambda=0.005$, $\Omega=1.0$, and $T=0.1$, using a reference unit of energy of 208.5~cm$^{-1}$ (300~K), which corresponds to the adiabatic regime under low temperatures. Also shown are numerically-exact results.}
\label{fig:intro}
\end{figure}

The purpose of this Paper is twofold. First, it demonstrates that the aforementioned inconsistencies between populations and coherences may yield unphysical density matrices. This is exemplified in Fig.~\ref{fig:intro}, where a violation of positivity is observed in the local (diabatic) basis for a trimer (three level system) coupled to a harmonic environment. These observations underscore the importance of recent efforts aimed at finding consistent FSSH-like formulations of the entire density matrix \cite{Wang.JCPL.2015, Wang.JCP.2015, Martens.JCP.2015, Martens.JPCL.2016, Tempelaar.CSH.JCP.2018, Martens.JPCA.2019}, leading up to the second purpose of this Paper of singling out a recently proposed generalization of FSSH, called coherent fewest-switches surface hopping (C-FSSH) \cite{Tempelaar.CSH.JCP.2018}. C-FSSH uniquely provides a consistent formulation of the entire density matrix based on the existing (and practically unaltered) FSSH method by invoking pairs of active surfaces. Meanwhile, the classical coordinates are represented by Gaussian wavepackets, allowing zero-point fluctuations to be represented without invoking a Wigner distribution. In addition to resolving the positivity violations in FSSH, we demonstrate C-FSSH to produce comparatively superior results throughout much (but not all) of parameter space for a trimeric system.

The remainder of this Paper is organized as follows. In Sec.~\ref{sec:properties}, 
we provide a summary of the physical properties of density matrices. In Sec.~\ref{sec:hamilt}, 
we describe some general concepts of mixed quantum--classical dynamics, while in Sec.~\ref{sec:FSSH} and Sec.~\ref{sec:CFSSH}, 
we introduce method details specific to FSSH and C-FSSH, respectively. 
In Sec.~\ref{sec:model}, we introduce the model system used in this study, 
after which we discuss the results in Sec.~\ref{sec:results}. 
Finally, we present our conclusions and outlook in Sec.~\ref{sec:outlook}.

\section{Physical properties of density matrices}
\label{sec:properties}

Before evaluating density matrices within the FSSH and C-FSSH formalisms, we begin by iterating the physical properties a valid (reduced) density matrix should satisfy. First, it should be noted that the density matrix of a pure state in some basis is given by $\rho_{nm}=c_n^*c_m^{\vphantom{^*}}$, with $c_n$ as the wavefunction coefficient in the basis at hand (for which basis states are labeled by $n$). For a mixed state, the density matrix instead takes the form $\rho_{nm}=\braket{c_n^\dagger c_m^{\vphantom{^\dagger}}}$, where $\braket{\ldots}$ denotes an average over pure states. The physical properties emanating from these principles are as follows.

(1) Hermitian property: The density matrix should be hermitian, $\rho^{\dagger}=\rho$. It should be noted that the Hermitian property survives a unitary transformation of $\rho$, meaning it carries over from one basis to another.

(2) Normalization: Populations represent probabilities, and therefore their sum should add up to one. As a result, the trace of the density matrix should amount to unity, $\trace ( \rho ) = 1$. Like the Hermitian property, normalization survives a unitary transformation of $\rho$.

(3) Positivity: Positivity of a density matrix can be evaluated in various forms, as outlined in the following.

(3a) Positive populations: As populations represent probabilities, the values they assume should be strictly positive, $\rho_{nn}\geq0$. However, this positivity criterion does not necessarily survive a unitary transformation, meaning that positivity in one basis does not imply positivity in another basis. This is the result of mixing of populations and coherences by the basis transformation. Hence, positivity in \emph{all possible bases} requires these elements of $\rho$ to be consistent, leading up to the next two properties.

(3b) Cauchy--Schwartz (CS) inequality: The CS inequality puts bounds on the off-diagonal elements of the density matrix in terms of the diagonal elements, as
\begin{equation}
	\rho_{nn}\rho_{mm} \geq \rho_{nm}\rho_{mn}.
\end{equation}
The CS inequality is saturated for pure states, i.e.,
$ \rho_{nn}\rho_{mm} = \rho_{nm}\rho_{mn} $, which is the result of the diagonal and off-diagonal elements being interrelated through $\rho_{nm}=c_n^*c_m^{\vphantom{^*}}$. For mixed states, on the other hand, it takes the form $ \rho_{nn}\rho_{mm} \geq \rho_{nm}\rho_{mn} $, since taking the average $\braket{\ldots}$ may yield destructive interference for off-diagonal elements, but not for diagonal elements. When a 2$\times2$ density matrix satisfies the CS inequality, positivity is ensured in \emph{any} basis, rendering the CS inequality a more complete requirement than positivity in a given basis (Property 3a). However, density matrices of size $3\times3$ and larger involve multiple CS inequalities (invoking pairs of basis states) that become basis dependent. The importance of having a basis-independent metric of positivity leads up to the next and last property.

(3c) Positive eigenvalues: Eigenvalues of a matrix represent the extrema of diagonal elements as a function of all possible unitary rotation angles. Hence, the extrema for the density matrix can be found by introducing the density operator,
\begin{equation}
    \hat{\rho}=\sum_{m,n}\rho_{m,n}\ket{m}\bra{n},
\end{equation}
and solving the eigenvalue equation $\hat{\rho}\ket{\alpha}=r_\alpha\ket{\alpha}$. Positivity of the density matrix then implies that $r_\alpha\geq0$ for all $\alpha$. This forms an unambiguous and basis-independent definition of positivity, regardless of the size of the density matrix.

\section{Theory and methods}\label{sec:methods}

Having summarized the physical properties a density matrix should satisfy, we now proceed to evaluate density matrices within the FSSH and C-FSSH formalisms, after first introducing the general concept of mixed quantum--classical dynamics.

\subsection{Mixed quantum--classical dynamics}
\label{sec:hamilt}

Mixed quantum--classical dynamics approximates the environment as classical, yielding a total Hamiltonian given by
	\begin{equation}
	\hat{H}( \boldsymbol{p}, \boldsymbol{q} ) = \hat{H}_\mathrm{q} + \hat{H}_\mathrm{q-c} (\boldsymbol{q}) + 
																H_\mathrm{c} ( \boldsymbol{p}, \boldsymbol{q} ),
																\label{eq:H_MQC}
	\end{equation}
where $\boldsymbol{q}$ and $\boldsymbol{p}$ are the classical position and momentum vectors, respectively. In Eq.~\ref{eq:H_MQC} we differentiate contributions from the quantum system (q) and the classical environment (c), as well as their mutual interaction (q--c), the latter two depending parametrically on the classical coordinates. From here onward, this dependence will be omitted in order to simplify the notation.

A mixed quantum--classical simulation amounts to propagating a swarm of classical trajectories, while quantities of interest are evaluated as a trajectory average. (For simplicity, we will use the same notation for trajectory average as for an average over pure states, i.e., $\braket{\ldots}$, even though these two averages are not strictly interchangeable.) Within each trajectory, the classical coordinates evolve according to Hamilton's equations using the Hamiltonian $\hat{H}_\mathrm{q-c} (\boldsymbol{q}) + H_\mathrm{c} (\boldsymbol{p},\boldsymbol{q} )$. Appropriately taking the expectation value of $\hat{H}_\mathrm{q-c} (\boldsymbol{q})$ in order to arrive at a meaningful classical potential energy is a source of ambiguity, leading to the many variants by means of which mixed quantum--classical dynamics can be implemented.

In concert with the classical dynamics, the quantum wavefunction is propagated through the time-dependent Schr{\"o}dinger equation (setting $\hbar=1$),
	\begin{equation}
	i \ket{\dot{\Psi}} = (\hat{H}_\mathrm{q}+\hat{H}_\mathrm{q-c}) \ket{\Psi}.
	\end{equation}
At any given time, the quantum wavefunction can be expanded in the adiabatic basis as
	\begin{equation}
	\ket{\Psi} = \sum_k c_k \ket{k},
	\label{eq:expansion}
	\end{equation}
with the adiabatic basis states following from the instantaneous eigensolution of the time-independent Schr{\"o}dinger equation,
	\begin{equation}
	(\hat{H}_\mathrm{q} + \hat{H}_\mathrm{q-c})\ket{k} = \epsilon_k \ket{k}.
	\end{equation}
Here, $\epsilon_k$ is the associated (adiabatic) energy.

While it may serve to intuitively assess quantities of interest, the adiabatic basis strictly does not allow for a trajectory average to be taken, as the basis itself is trajectory dependent. Rather, a diabatic (trajectory independent) basis needs to be adopted, such as the local basis, within which the adiabatic states are expanded as
	\begin{equation}
	\ket{k} = \sum_n U_{k,n} \ket{n},
	\end{equation}
where the unitary transformation $U_{k,n}$ itself is trajectory dependent. Here and henceforth, we use $k$ and $l$ to denote adiabatic states, and $n$ and $m$ to denote local basis states. Moreover, as much as possible we will use the local basis to assess whether a given density matrix satisfies the physical properties summarized in Sec.~\ref{sec:properties}.

\subsection{Fewest-switches surface hopping (FSSH)}
\label{sec:FSSH}

We proceed with a quick summary of FSSH, referring to other work for a more elaborate explanation \cite{Hammes.JCP.1994}. FSSH is an example of a broader class of surface hopping methods, wherein the classical potential energy is derived from an expectation value of the quantum--classical Hamiltonian with respect to a single adiabatic surface, i.e., $\braket{a|\hat{H}_\mathrm{q-c}|a}$. FSSH allows for stochastic switches between surfaces to happen at any time, with the probability of switching from surface $k$ to $l$ being governed by the probability
\begin{equation}
	P_{a:k \rightarrow l} = 2 \mathrm{Re} \Big(  \boldsymbol{p} \cdot 
						\boldsymbol{d}_{k,l} \frac{c_l}{c_k} \Delta t   \Big).
\end{equation}
Here, $c_k$ is the wavefunction expansion coefficient from Eq.~\ref{eq:expansion}, $\Delta t$ is the time integration step, and $\boldsymbol{d}_{k,l}$ is the nonadiabatic coupling vector defined as
\begin{equation}
	\boldsymbol{d}_{k,l} \equiv \braket{k | \nabla_q | l}.
\end{equation}
In order to limit the number of switches, a uniform random number $ \xi \in (0,1)$ is drawn at each time step, and the switch from $k$ to $l$ is allowed to occur only when
\begin{equation}
	\sum_{l' \leq l-1}  P_{a:k \rightarrow l'} < \xi <  \sum_{l' \leq l}  P_{a:k \rightarrow l'} .
\end{equation}
Moreover, in order to conserve (quantum plus classical) energy, the change in quantum energy upon a switch, $\epsilon_k - \epsilon_l$, needs to be absorbed or donated in the form of an adjustment of the classical kinetic energy, $\mathbf{p}^2/2$. Accordingly, the classical momentum is rescaled in the direction of the nonadiabatic coupling vector. Whenever there is insufficient classical kinetic energy available the switching event is abandoned.

As mentioned in the Introduction, FSSH was originally developed with the intention to study scattering phenomena. In such cases, the active surface can be initiated in some adiabatic state as $a=i$ and scattering probabilities can be assessed by taking a trajectory average of $a$ at a later time. As pointed out in Sec.~\ref{sec:hamilt}, such a trajectory average within the adiabatic basis is not strictly allowed, yet it provides a good approximation in the asymptotic regimes of scattering phenomena where the adiabatic basis becomes near-diabatic. A more rigorous approach to quantify scattering outcomes, however, is to consider the density matrix in the local basis \cite{Muller.JCP.1997, Kelly.JCP.2013, Landry.JCP.2013},
\begin{equation}
	\rho_{n,m} = \Braket{\sum_{k,l} U_{k,n} \rho_{k,l}  U_{l,m}^*},
	\label{eq:DM_original_diab}
\end{equation}
with the trajectory-dependent adiabatic contributions given by \cite{Muller.JCP.1997, Kelly.JCP.2013, Landry.JCP.2013}
\begin{equation}
	\rho_{k,l} = \delta_{k,l} \delta_{k,a}.
	\label{eq:DM_original}
\end{equation}
In order to assess whether $\rho_{n,m}$ satisfies the physical properties summarized in Sec.~\ref{sec:properties}, it is worth pointing out that the adiabatic contributions given by Eq.~\ref{eq:DM_original} each trivially satisfy these properties, which carries over to $\rho_{n,m}$.

Where Eqs.~\ref{eq:DM_original_diab} and \ref{eq:DM_original} fall short is in their inability to account for a sharing of initial quantum population among different surfaces. This can be straightforwardly incorporated by taking $\rho_{n,m} = \braket{\sum_i \rho_{n,m}^{(i)}}$, with $\rho_{n,m}^{(i)} = \sum_{k,l} U_{k,n}^{(i)} \rho_{k,l}^{(i)} {U_{l,m}^{(i)}}^*$ and with
\begin{equation}
	\rho_{k,l}^{(i)} = \rho_0^{(i)} \delta_{k,l} \delta_{k,a^{(i)}}.
	\label{eq:DM_incoh}
\end{equation}
Here, $i$ refers to a ``branch'' of the trajectory that was initiated in surface $i$, meaning that $a^{(i)}=i$ initially. In addition to the active surface, this branch comes with its own set of classical coordinates $\boldsymbol{p}^{(i)}$ and $\boldsymbol{q}^{(i)}$ as well as wavefunction coefficients $c_k^{(i)}$, each initiated using the same (branch-independent) value. Moreover, each branch is weighted by $\rho_0^{(i)}$ in Eq.~\ref{eq:DM_incoh}. If $\sum_i\rho_0^{(i)}=1$ is enforced, it follows that $\sum_n\rho_{n,n}=1$ and normalization is satisfied. Likewise, $\rho_{n,m}$ can be shown to satisfy the other physical properties summarized in Sec.~\ref{sec:properties}.

The adiabatic contribution to the density matrix embodied by Eq.~\ref{eq:DM_incoh} is still diagonal, and as such completely neglects adiabatic coherences. Such coherences can be incorporated by supplementing Eq.~\ref{eq:DM_incoh} with off-diagonal elements based on the wavefunction coefficients, as \cite{Tempelaar.JCP.2013, Landry.JCP.2013, Tempelaar.mapping.JPCL.2014, vanderVegte.JPCB.2014, vanderVegte.JPCB.2015,  Chen.JCP.2016,  Peng.JCP.2019}
\begin{equation}
	\rho_{k,l}^{(i)} = \rho_0^{(i)} \big( \delta_{k,l} \delta_{k,a^{(i)}} + (1 - \delta_{k,l}) c_k^{(i)} c_l^{(i)*} \big).
	\label{eq:DM_coh}
\end{equation}
It can be easily checked that this supplement does not contribute to $\sum_n\rho_{n,n}$ so that normalization is satisfied as long as $\sum_i\rho_0^{(i)}=1$. Moreover, if the wavefunction coefficients are initiated such that $\rho_0^{(i)}=\vert c_i\vert^2$, it can be shown that initially $\rho_{n,m}=c_n^*c_m^{\vphantom{^*}}$ as a result of which the physical properties outlined in Sec.~\ref{sec:properties} are automatically satisfied. However, as time evolves, the diagonal and off-diagonal elements start to behave markedly different, being constructed out of active surfaces and wavefunction coefficients, respectively. As a result of such inconsistencies, positivity may become violated, which is most intuitively understood as a consequence of the inconsistencies breaking the CS inequality (Property 3b). As a result, negative populations may arise in some basis, as found in Fig.~\ref{fig:intro}.

\subsection{Coherent fewest-switches surface hopping (C-FSSH)}
\label{sec:CFSSH}

If one propagates multiple branches for each trajectory as outlined in Sec.~\ref{sec:FSSH}, one can \emph{in principle} construct off-diagonal elements of the density matrix based on pairs of branches. This is the core idea behind C-FSSH, which has been described in detail in Ref.~\citenum{Tempelaar.CSH.JCP.2018}, and for which the density matrix is constructed as $\rho_{n,m} = \braket{\sum_{i,j} \rho_{n,m}^{(i,j)}}$. Here, $\rho_{n,m}^{(i,j)} = \sum_{k,l} U_{k,n}^{(i,j)} \rho_{k,l}^{(i,j)} {U_{l,m}^{(i,j)}}^*$ with
	\begin{equation}
	\rho_{k,l}^{(i,j)} = \rho_0^{(i,j)}  \delta_{k,a^{(i)}} \delta_{l,a^{(j)}}  F^{(i,j)}   e^{-i ( \phi_k^{(i)} - \phi_l^{(j)} )},
	\label{eq:DM_CFSSH}
	\end{equation}
where $(i,j)$ labels the pair of branches, the contribution of which is weighted by $\rho_0^{(i,j)}$. When this weight is derived from the initial wavefunction coefficients as $\rho_0^{(i,j)}=c_i^*c_j^{\vphantom{^*}}$, it can again be shown that initially $\rho_{n,m}=c_n^*c_m^{\vphantom{^*}}$. Interestingly, the branch-dependent wavefunction coefficients do not need to be uniformly initiated (as $c_k^{(i)}=c_k^{\vphantom{^(i)}}$), and it has been found that $c_k^{(i)}=\delta_{k,i}$ yields improved accuracy \cite{Tempelaar.CSH.JCP.2018}. In addition to the wavefunction coefficients and the active surface, initiated as $a^{(i)}=i$, each branch features a phase factor that is propagated as $\dot{\phi}_k^{(i)}=\epsilon_k^{(i)}$. Here, $\epsilon_k^{(i)}$ is the $k$th eigenvalue of the quantum Hamiltonian $\hat{H}_\mathrm{q}+\hat{H}_\mathrm{q-c}^{(i)}$, which depends parametrically on the branch-dependent classical coordinates $\boldsymbol{q}^{(i)}$ through $\hat{H}_\mathrm{q-c}^{(i)}$. One can also define a Hamiltonian for each branch pair, $\hat{H}_\mathrm{q}+\hat{H}_\mathrm{q-c}^{(i,j)}$, from which the unitary transformation $U_{k,n}^{(i,j)}$ is derived. This Hamiltonian depends parametrically on branch pair-dependent classical coordinates, which can simply be taken to be $\boldsymbol{q}^{(i,j)}=(\boldsymbol{q}^{(i)}+\boldsymbol{q}^{(j)})/2$.

The last term appearing in Eq.~\ref{eq:DM_CFSSH} represents a classical overlap factor. Accordingly, the classical coordinates are interpreted as describing the centers of wavepackets with a finite spread in position and momentum space. This allows incorporation of the Heisenberg uncertainty principle within the classical environment and, consequently, zero-point fluctuations. When constructing the reduced density matrix of the quantum system, by tracing out the classical coordinates, the classical wavepackets contribute an overlap factor of the form \footnote{We note that an absolute value was taken in the previous work introducing C-FSSH \cite{Tempelaar.CSH.JCP.2018} based on empirical grounds, although it can be argued that this was not physically motivated. Such has been omitted here, bearing in mind that the final form of the overlap factor is real and positive by construction.}
\begin{equation}
	F^{(i,j)} =  \braket{\boldsymbol{p}^{(i)}, \boldsymbol{q}^{(i)} | \boldsymbol{p}^{(j)}, \boldsymbol{q}^{(j)}}
\end{equation}
with
\begin{align}
	\braket{\boldsymbol{p}^{(i)}, \boldsymbol{q}^{(i)} | \boldsymbol{p}^{(j)}, \boldsymbol{q}^{(j)}} 
	=& 
	\prod_{\alpha}  \exp \Big( - \frac{\sigma_q^2}{4} 
	( p_{\alpha}^{(i)} - p_{\alpha}^{(j)} )^2  \Big) \nonumber \\
	&\times
	\exp \Big( - \frac{1}{4 \sigma_q^2} ( q_{\alpha}^{(i)} - q_{\alpha}^{(j)} )^2  \Big) \label{eq:overlap_complete} \\
	&\times
	\exp \Big( \frac{i}{2} ( p_{\alpha}^{(i)} + p_{\alpha}^{(j)} )  
	( q_{\alpha}^{(i)} - q_{\alpha}^{(j)} )  \Big). \nonumber
\end{align}
Here, $ \sigma_q $ represents the width of the classical wavepackets. Upon introducing C-FSSH \cite{Tempelaar.CSH.JCP.2018} this width was taken to be a constant, consistent with the concept of frozen Gaussians introduced by Heller \cite{Heller.JCP.1981}. However, whereas Heller's approach shaped the Gaussian after the original wavepacket, $ \sigma_q $ was effectively treated as an arbitrary parameter in C-FSSH, following previous examples in the literature \cite{Kluk.JCP.1986, Thompson.ChemPhys.2010}. In the following we demonstrate that this parameter can be eliminated by maximizing the overlap (thus minimizing the argument in Eq.~\ref{eq:overlap_complete}), yielding
\begin{equation}
	\braket{\boldsymbol{p}^{(i)}, \boldsymbol{q}^{(i)} | \boldsymbol{p}^{(j)}, \boldsymbol{q}^{(j)}} 
	= 
	\prod_{\alpha}  \exp \Big( -\frac{1}{2} | ( p_\alpha^{(i)} - p_\alpha^{(j)} ) 
	( q_\alpha^{(i)} - q_\alpha^{(j)} ) |  \Big).
\end{equation}
The resulting parameter-free treatment of the classical overlap is consistently adopted in the present Paper, and based on the favorable results it generates (\emph{vide infra}) we conclude that this treatment forms an improvement over the original C-FSSH method.

As before, in order to assess whether the density matrix satisfies the physical properties summarized in Sec.~\ref{sec:properties}, one can examine the adiabatic contributions, which are now trajectory \emph{and branch-pair} dependent. First, it can easily be verified that Eq.~\ref{eq:DM_CFSSH} by construction satisfies Property 3b, i.e., the CS inequality,
	\begin{equation}
    \rho_{k,k}^{(i,j)}\rho_{l,l}^{(i,j)}\geq \rho_{k,l}^{(i,j)}\rho_{l,k}^{(i,j)},
    \label{eq:CS_CFSSH}
    \end{equation}
which is a result of the preserved consistency between diagonal and off-diagonal elements. However, it can also easily be verified that Eq.~\ref{eq:DM_CFSSH} does not strictly satisfy normalization, since a switching resulting in $a^{(i)}=j$ within branch $(i,j)$ yields an excess contribution to the trace. Although this can be remedied by renormalizing the density matrix, for C-FSSH it was instead suggested to decouple each off-diagonal density matrix elements from \emph{all other} elements, such that for $k\neq l$ only the branch pair with $i=k$ and $j=l$ contributes,
	\begin{equation}
	\rho_{k,l}^{(k,l)} = \rho_0^{(k,l)}  \delta_{k,a^{(k)}} \delta_{l,a^{(l)}}  F^{(k,l)} e^{-i ( \phi_k^{(k)} - \phi_l^{(l)} )}.
	\label{eq:CFSSH_offdiag}
	\end{equation}
This leaves diagonal density matrix elements of the form
\begin{equation}
	\rho_{k,k}^{(i,i)} = \rho_0^{(i,i)} \delta_{k,a^{(i)}},
\end{equation}
upon which a conservation of the trace is restored. As a downside of this decoupling, it can no longer be trivially shown that the CS inequality is automatically satisfied, as Eq.~\ref{eq:CS_CFSSH} can no longer be formulated. What can be shown, however, is that at instances where all branches happen to share the same classical coordinates, $\boldsymbol{q}^{(i,j)}=\boldsymbol{q}$, meaning there is a well-defined, branch-pair-independent adiabatic basis, we have
\begin{equation}
	\sum_{i,j}\rho_{k,k}^{(i,j)}\rho_{l,l}^{(i,j)}\geq \sum_{i,j}\rho_{k,l}^{(i,j)}\rho_{l,k}^{(i,j)}.
	\label{eq:CS_rigorous}
\end{equation}
Although rather restrictive, this example suggests that the CS inequality would have to be satisfied at least approximately in the more general case. For a less ambiguous demonstration, however, numerical calculations are necessary, as presented in the next Section. This also applies to Property 3c, for which no analytical relationship can generally be given.

\section{Application to a trimer}
\label{sec:trimer}

In the previous Section, we contributed analytical arguments of why unphysical density matrices arise in FSSH, but not in C-FSSH. In the present Section, we substantiate these arguments with numerical results for a trimer system. We should point out that no pronounced unphysical behaviors were observed in dimeric systems (not shown here, but for a comprehensive characterization we refer to Ref.~\citenum{Tempelaar.CSH.JCP.2018}). It therefore seems that the combination of multiple pairwise interactions between quantum levels is necessary for such behaviors to become prevalent. In addition to being the simplest system incorporating multiple pairwise interactions, a trimer is an important model for donor--bridge--acceptor systems widely used for energy and spin transfer processes \cite{Albinsson.JPPC.2006, Sifain.JCP.2019, Valianti.JPCL.2022}.

\subsection{Model}
\label{sec:model}

The trimer model considered here is effectively a three-site tight-binding model with open boundary conditions, such that interactions are limited to nearest-neighbors. Within the local basis, the Hamiltonian is given by
	\begin{equation}
	  \hat{H}_\mathrm{q} = 
	  \begin{pmatrix}
	    E_{1} & V & 0 \\
	    V & E_{2} & V \\
	    0 & V & E_{3} 
	  \end{pmatrix},
	\end{equation}
where $V$ is the interaction strength and $E_n$ is the diagonal energy associated with local basis state $n$.

The classical coordinates are represented by harmonic oscillators, as detailed in Ref.~\citenum{Tempelaar.CSH.JCP.2018}. Bilinear coupling between these coordinates and the quantum system is incorporated through the quantum--classical interaction Hamiltonian
	\begin{equation}
	  \hat{H}_\mathrm{q-c} = 
	  \begin{pmatrix}
	    K_{1} & 0 & 0 \\
	    0 & K_{2} & 0 \\
	    0 & 0 & K_{3} 
	  \end{pmatrix},
	\end{equation}
with $K_n\equiv\sum_\alpha g_{n,\alpha}q_\alpha$, and where $g_{n,\alpha}$ is the coupling constant associated with quantum level $n$ and classical coordinate $q_\alpha$. We assume that each coordinate couples exclusively to a single level $n$, so that the different $K_n$ are fully uncorrelated. The coupling constants are determined based on a Drude--Lorentz spectral density (also known as, 
the overdamped Brownian oscillator model \cite{Tanimura.stochastic.JPhysSocJpn.2006, 
Mukamel.1995.B01}), taken to be identical for each $n$,
	\begin{equation}
	J_n(\omega) \equiv \frac{\pi}{2}\sum_\alpha\frac{g_{n,\alpha}}{\omega_\alpha}\delta(\omega-\omega_\alpha) = 2 \lambda  \frac{\omega \Omega}{\omega^2 + \Omega^2},
	\end{equation}
where $\omega_\alpha$ is the frequency associated with classical coordinate $q_\alpha$, and where $\Omega$ and $\lambda$ are the spectral density characteristic frequency and reorganization energy, respectively. It should be stressed, however, that mixed quantum--classical methods are generally capable of treating arbitrary spectral densities as well as anharmonic effects.

For a given temperature $T$, the initial classical coordinates $\boldsymbol{p}_0$ and $\boldsymbol{q}_0$ can be sampled from a Boltzmann distribution
	\begin{equation}
	P( \boldsymbol{p}_0, \boldsymbol{q}_0 ) \propto \prod_{\alpha} \exp 
	\Big[  -\beta  \Big(  \frac{1}{2} p_{0,\alpha}^2 + 
	\frac{1}{2} \omega_{\alpha}^2 q_{0,\alpha}^2   \Big)      \Big],
	\label{eq:Boltz_dist}
	\end{equation}
where $\beta=1/T$ and the Boltzmann constant and classical masses are taken to be unity. For FSSH, this distribution can be replaced by the Wigner distribution,
	\begin{equation}
	P( \boldsymbol{p}_0, \boldsymbol{q}_0 ) \propto \prod_{\alpha} \exp 
	\Big[  - \frac{2}{\omega_{\alpha}} \tanh  \Big( \frac{\beta \omega_{\alpha}}{2}  \Big) 
	\Big(  \frac{1}{2} p_{0,\alpha}^2 + 
	\frac{1}{2} \omega_{\alpha}^2 q_{0,\alpha}^2   \Big)      \Big],
	\label{eq:Wign_dist}
	\end{equation}
in order to account for zero-point fluctuations that may become particularly significant at low temperatures. For C-FSSH, however, such fluctuations are already represented by the classical wavepackets, such that no Wigner sampling is necessary (or desirable).

\subsection{Results}
\label{sec:results}

For a systematic survey of the trimer system we refer to the Supplementary Material, where results from C-FSSH are compared against those from FSSH under Boltzmann sampling [FSSH (B)] and Wigner sampling [FSSH (W)] (through Eqs. \ref{eq:Boltz_dist} and \ref{eq:Wign_dist}, respectively) as well as numerically-exact results from the hierarchical equation of motion \cite{Tanimura.1989} (as implemented in the Parallel Hierarchy Integrator
\cite{Strumpfer.JCTC.2012}). Here, a high-dimensional sweep over parameter space is performed including variations in $V$, $\lambda$, $\Omega$, and $T$. Moreover, three different trimer systems are considered, characterized by the diagonal energies $E_n$, the first being a homogeneous trimer with $E_1=E_2=E_3$, the second being a biased trimer with $E_1-E_2=E_2-E_3=0.5$, and the third being a donor--bridge--acceptor system with $E_2-E_1=E_1-E_3=1.0$, where the reference unit of energy is taken to be 208.5~cm$^{-1}$ (which is the thermal quantum at 300~K). In all cases, the quantum system is initiated in the first level, $n=1$. In the following, we will be discussing select cases of interest.

\begin{figure}[t]
\centering
\includegraphics{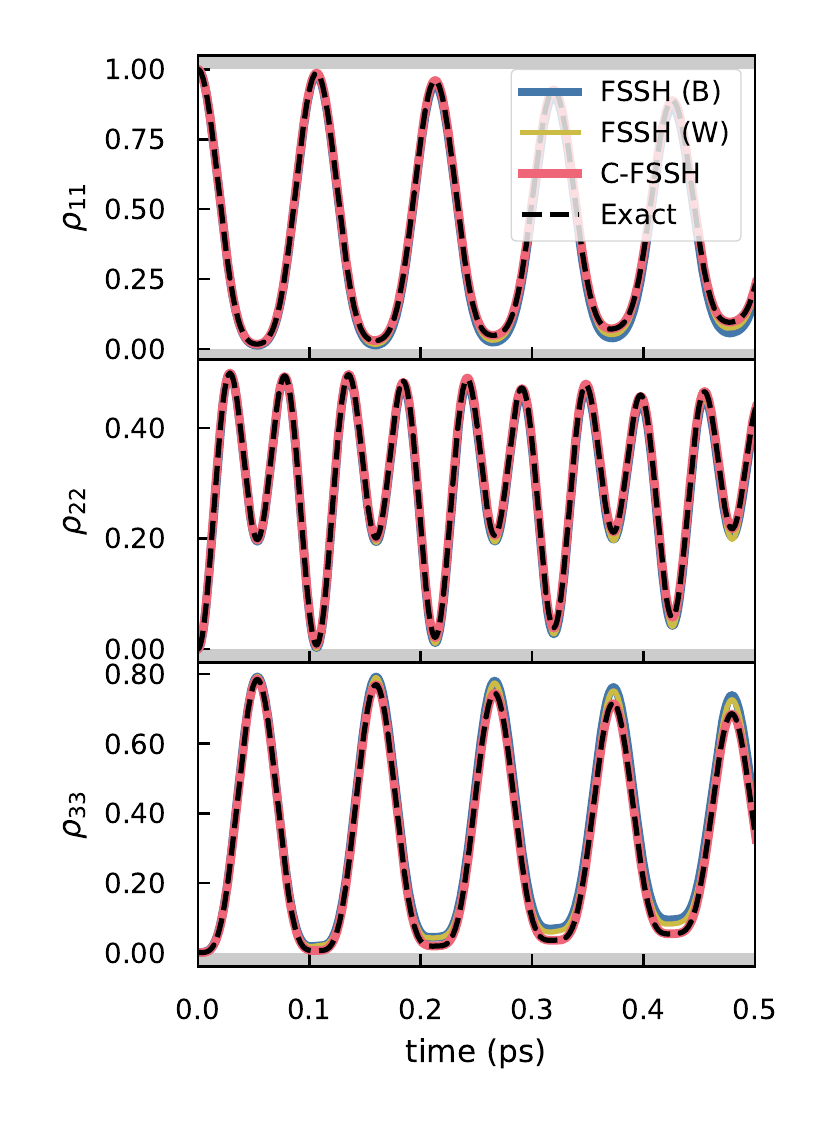}

\caption{Transient population of a biased trimer in the local basis. Region of unphysical values is highlighted in grey for ease of demonstration. Applied parameter are $V=1.0$, $\lambda=0.005$, $\Omega=0.1$, and $T=1.0$, using a reference unit of energy of 208.5~cm$^{-1}$ (300~K), which corresponds to the adiabatic regime under high temperatures. Numerically-exact results are shown alongside the results from C-FSSH and FSSH under Boltzmann [FSSH (B)] and Wigner sampling of classical coordinates [FSSH (W)].
}
\label{fig:pop_diab_HT}
\end{figure}

The result shown in Fig.~\ref{fig:intro} was obtained using FSSH (W) for a homogeneous trimer with $V=1.0$, $\lambda=0.005$, $\Omega=1.0$, and $T=0.1$, which corresponds to adiabatic regime at low temperature. It should be pointed out that the low temperature regime, where $\Omega > T$, is notoriously demanding for mixed quantum--classical methods, although a previous survey for the spin-boson model has shown C-FSSH to reach surprising accuracy in this regime \cite{Tempelaar.CSH.JCP.2018}. Regardless, it is of interest to assess whether violations of positivity occur in the high temperature regime where instead $\Omega < T$, and where mixed quantum--classical dynamics is generally expected to behave well. To this end we show in Fig.~\ref{fig:pop_diab_HT} the complete set of local populations for a biased trimer with $V=1.0$, $\lambda=0.005$, $\Omega=0.1$, and $T=1.0$. Results are shown for FSSH (B), FSSH (W), as well as C-FSSH, alongside numerically-exact results.

\begin{figure}[t]
\centering
\includegraphics{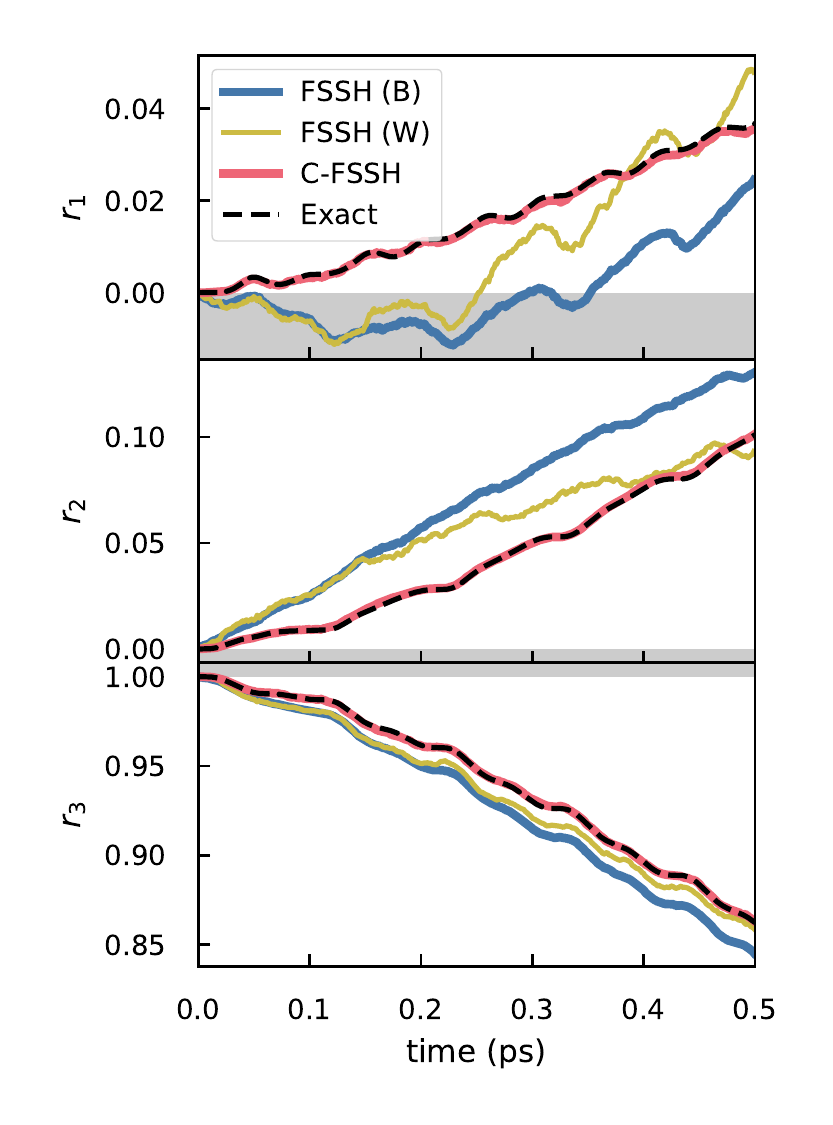}

\caption{Same as Fig.~\ref{fig:pop_diab_HT}, but for eigenvalues of the density matrix. 
}
\label{fig:EigVal_diab_HT}
\end{figure}

Interestingly, in Fig.~\ref{fig:pop_diab_HT}, C-FSSH is shown to yield an improved accuracy compared to FSSH (B) and FSSH (W), but none of the methods predict negative populations in the local basis. However, as pointed out in Sec.~\ref{sec:properties}, this does not necessarily mean that positivity is preserved in \emph{any} basis. To assess this, one should instead consider Property 3c, that is, the eigenvalues of the density matrix, which are depicted in Fig.~\ref{fig:EigVal_diab_HT}. From this, violations of positivity are observed for both FSSH (B) and FSSH (W), as eigenvalues assume negative values.

\begin{figure}[htbp]
\centering
\includegraphics{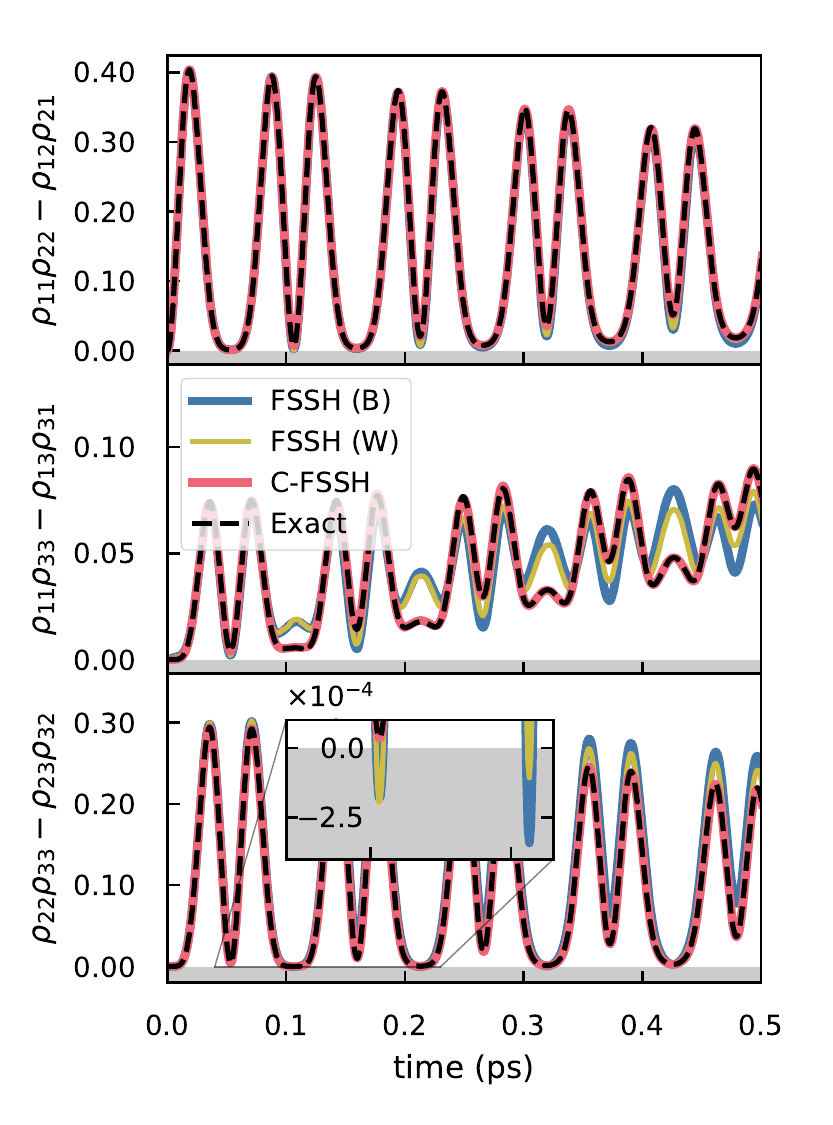}

\caption{Same as Fig. \ref{fig:pop_diab_HT}, but for CS inequalities.}
\label{fig:CSI_diab_HT}
\end{figure}

Whereas Property 3c is the most robust means to assess positivity violations, a more intuitive assessment is provided by Property 3b, i.e., the CS inequality, as it directly invokes the density matrix elements in some given basis. To this end, we plot in Fig.~\ref{fig:CSI_diab_HT} the quantity $\rho_{nn}\rho_{mm}-\rho_{nm}\rho_{mn}$ for each pair of states. Once this quantity becomes negative, we have a violation of the CS inequality, and thus a violation of positivity. From Fig.~\ref{fig:CSI_diab_HT}, FSSH (B) and FSSH (W) are seen to violate the CS inequality particularly for $n=2$ and $m=3$, although the degree at which this violation occurs is very small.

\begin{figure}[htbp]
\centering
\includegraphics{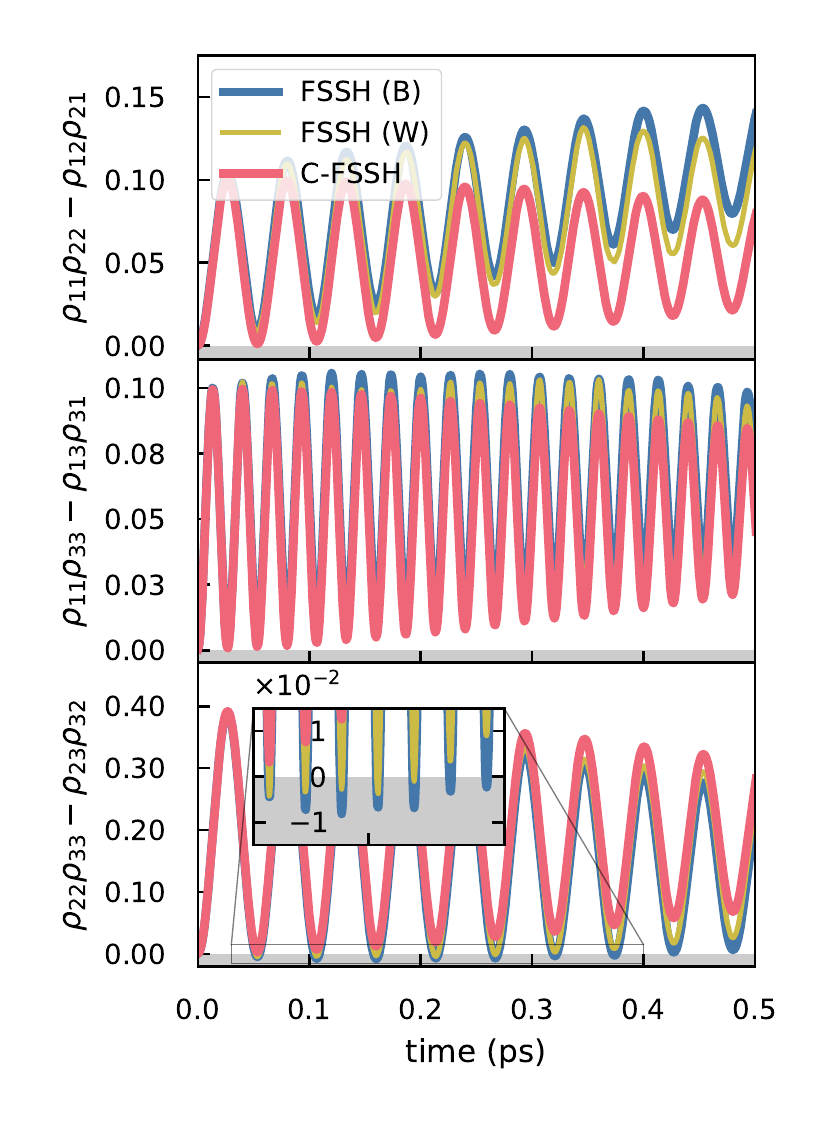}

\caption{Transient CS inequalities in the adiabatic basis (which is assured to be well-defined by restricting the classical coordinates to be identical among different branches/trajectories). Parameters are as in Fig.~\ref{fig:pop_diab_HT}.}
\label{fig:CSI_adiad_HT}
\end{figure}

An even more intuitive analysis of positivity is provided by the CS inequality in the adiabatic basis, as this is the basis in which analytical expressions for the involved density matrix elements are given; cf.~Eqs.~\ref{eq:DM_coh} and \ref{eq:DM_CFSSH}. As noted in Sec.~\ref{sec:hamilt}, however, such analysis is only meaningful when at a given time the classical coordinates are taken to be the same for every branch and every trajectory, so that a well-defined adiabatic basis exists. We have performed an additional simulation continuously reinforcing this condition by initiating all classical coordinates identically, and by neglecting the quantum contribution to the classical potential energy as well as by neglecting the adjustment of the classical kinetic energy upon a switch of the active surface. Under these constraints, C-FSSH rigorously satisfies the CS inequality, cf.~Eq.~\ref{eq:CS_rigorous}, which is borne out in Fig.~\ref{fig:CSI_adiad_HT} where the CS inequality is assessed for the same parameters as in Fig.~\ref{fig:CSI_diab_HT}. Here, FSSH (B) and FSSH (W) are seen to significantly violate the CS inequality as the product of wavefunction coefficients statistically exceeds the average contributions due to active surfaces. Overall, the CS inequalities behave markedly different in the adiabatic basis as compared to the local basis, underscoring the basis dependence of this property. It should be pointed out that the entanglement between the system and environment prohibits the extraction of numerically-exact results within the adiabatic basis, as a result of which no numerically-exact results are shown in Fig.~\ref{fig:CSI_adiad_HT}.

\begin{figure}[htbp]
\centering
\includegraphics{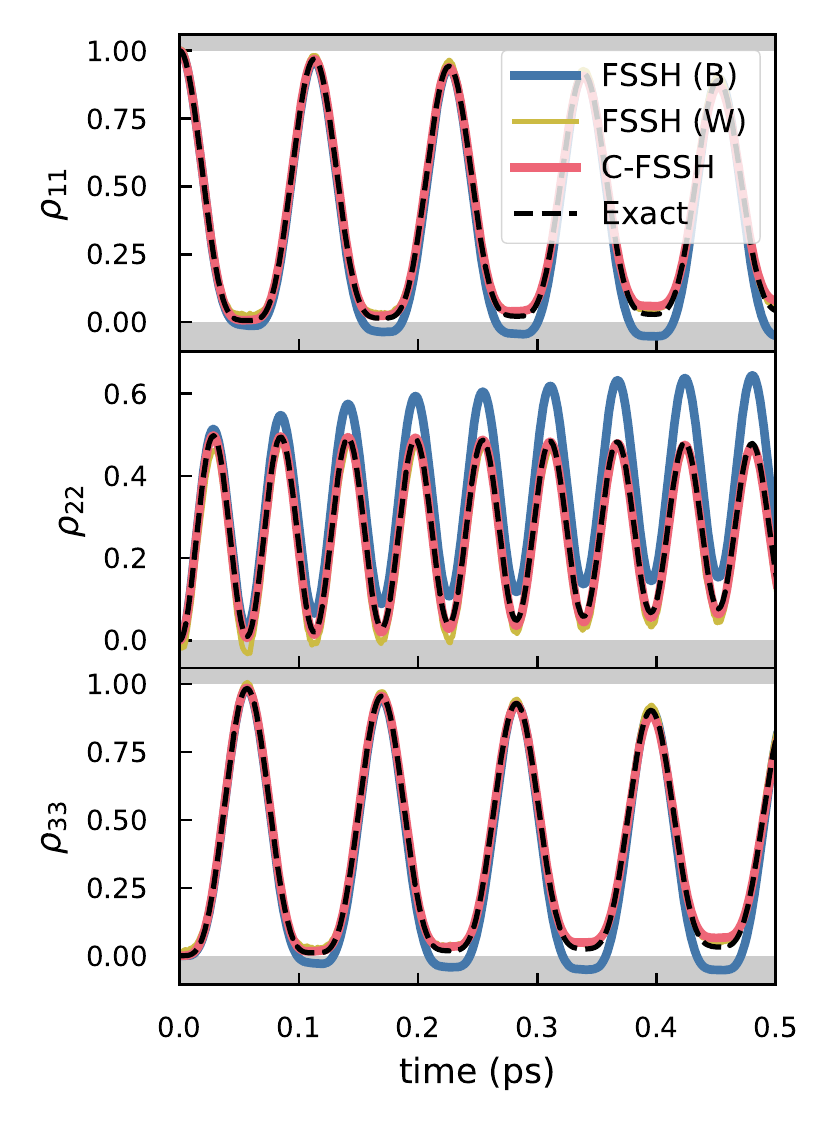}

\caption{Same as Fig.~\ref{fig:pop_diab_HT} but for a homogeneous trimer with $V=1.0$, $\lambda=0.025$, $\Omega=1.0$, and $T=0.1$, using a reference unit of energy of 208.5~cm$^{-1}$ (300~K). This corresponds to the adiabatic regime under low temperatures. }
\label{fig:pop_LT_0.025}
\end{figure}

\begin{figure}[htbp]
\centering
\includegraphics{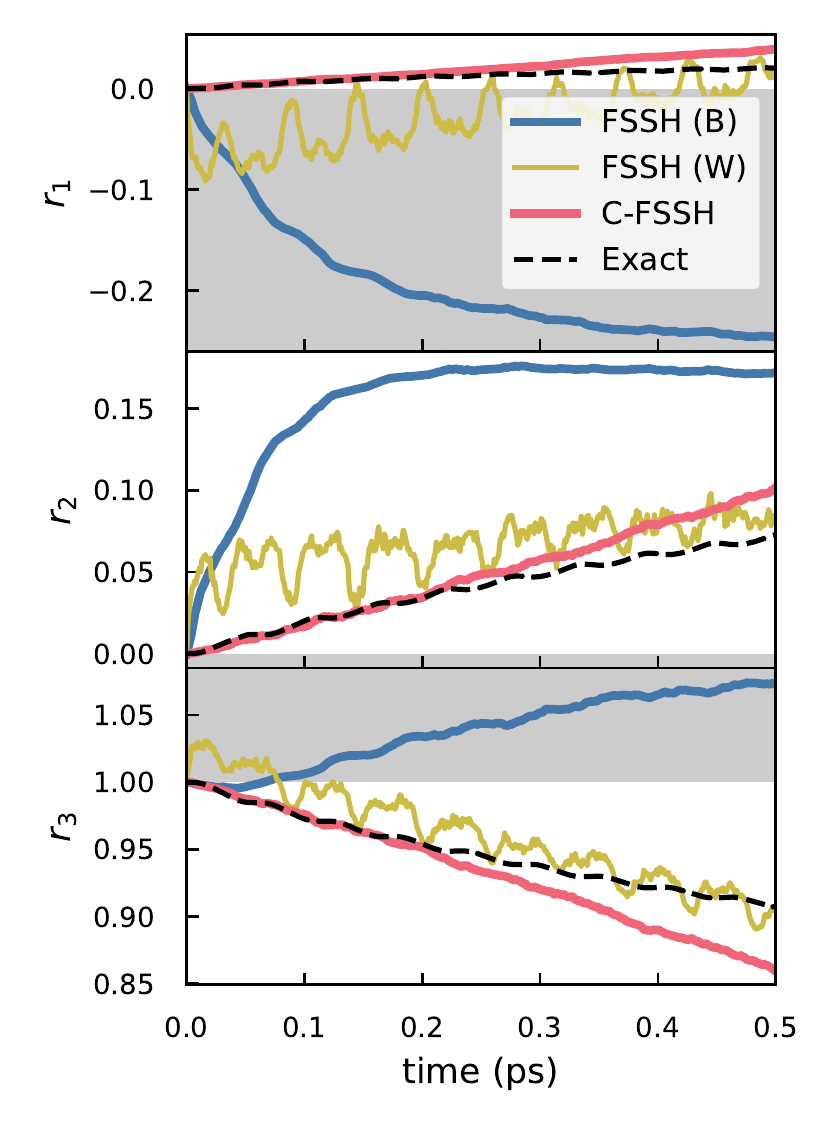}

\caption{Same as Fig.~\ref{fig:pop_LT_0.025}, but for eigenvalues of the density matrix.}
\label{fig:EigVal_LT_0.025}
\end{figure}

Lastly, we revisit the low-temperature regime, and present results for a homogeneous trimer with $V=1.0$, $\lambda = 0.025$, $\Omega=1.0$, and $T=0.1$. Shown in Figs.~\ref{fig:pop_LT_0.025} and \ref{fig:EigVal_LT_0.025} are local populations and eigenvalues of the density matrix, respectively. Significant violations of Property 3c are seen for FSSH (B) and FSSH (W), with eigenvalues being strongly negative. For FSSH (B) this leads to pronounced negative values of $\rho_{11}$ and $\rho_{33}$. Interestingly, FSSH (W) leads to an overall improvement in accuracy while enforcing positive values for $\rho_{11}$ and $\rho_{33}$, but yields negative values for $\rho_{22}$ instead. C-FSSH is once more seen to satisfy positivity throughout, while yielding superior accuracy.

C-FSSH consistently satisfies positivity throughout all of the parameter space covered in the systematic survey presented in the Supplementary Material, underscoring the robustness of this method in reinforcing the physical properties of the density matrix. FSSH (B) and FSSH (W), on the other hand, are seen to frequently violate positivity in the form of negative density matrix eigenvalues. At high temperatures, this is particularly prevalent for small reorganization energy, whereas at low temperatures violations are observed more consistently. It should be noted that imposing physical properties of the density matrix does not necessarily imply an improvement in accuracy. As reported before \cite{Tempelaar.CSH.JCP.2018}, C-FSSH is oftentimes seen to yield a radically-improved accuracy compared to FSSH (B and W), but not always, which is underscored by the survey presented in the Supplementary Material. In particular, FSSH oftentimes performs better in the Marcus regime where the interaction strength $V$ is weak.

\section{Conclusions and Outlook}
\label{sec:outlook}

With the growing interest in the application of FSSH to evaluate the quantum dynamics for an entire reduced density matrix, this method is increasingly prompted to satisfy the physical properties that alternative quantum dynamical methods have been subject to. In this Paper, we have demonstrated a violation of positivity for density matrices obtained within FSSH following the prevalent implementation \cite{Tempelaar.JCP.2013, Landry.JCP.2013} where adiabatic coherences are constructed based on wavefunction coefficients, rather than the active surfaces that determine populations. We have furthermore shown that C-FSSH \cite{Tempelaar.CSH.JCP.2018}, which invokes a density matrix constructed entirely out of active surfaces, does not suffer from such positivity violations. While a formal proof of positivity within C-FSSH is complicated by the secular approximation taken to conserve the trace of the density matrix, our numerical survey for a trimeric system revealed this property to be consistently satisfied.

In many cases, the reinforcement of positivity by C-FSSH yields an overall improvement in accuracy over FSSH when compared to numerically-exact results. This renders C-FSSH an attractive formalism, especially considering that it comes at virtually the same cost as FSSH. Moreover, as shown in the present Paper, it does not require the introduction of additional parameters. However, there are instances where FSSH provides better accuracy, especially so when interactions between quantum levels are weak, reorganization energies are large, and quantum levels are degenerate. In such cases, inaccuracies for C-FSSH are predominantly manifested as excessive coherence decay, likely caused by an overestimation of hopping rates that cause the adiabatic off-diagonal density matrix elements to become zero (cf.~Eq.~\ref{eq:CFSSH_offdiag}). Interestingly, overestimations of hopping rates within this very parameter regime have previously been shown to be resolvable by augmented FSSH (A-FSSH) \cite{Subotnik.JCP.2011, Subotnik.JPCA.2011, Landry.JCP.2012}, and it would therefore be interesting to combine A-FSSH with the coherent generalization embodied by C-FSSH with the aim to yield universal improvements in accuracy.

It would furthermore be of interest to assess positivity for other methods providing consistent formulations of the entire density matrix based on surface hopping approaches \cite{Wang.JCPL.2015, Wang.JCP.2015, Martens.JCP.2015, Martens.JPCL.2016, Tempelaar.CSH.JCP.2018, Martens.JPCA.2019}. Assuring physical and consistent density matrices based on surface hopping are of particular importance for applications to spectral simulations, as well as phenomena that lie markedly outside the Marcus regime. As such, the present work may also be of interest to the recently-proposed reciprocal-space formulation \cite{Krotz.JCP.2011} of FSSH \cite{Krotz.JCP.2022} for the modeling of bandlike phenomena.

\begin{acknowledgments}
This material is based upon work supported by the National Science Foundation under Grant No.~2145433.
\end{acknowledgments}

\bibliography{ankabib_extracted}

\end{document}


\maketitle


\newpage

\begin{figure}[htbp]
\centering

\includegraphics[width=1.0\textwidth]{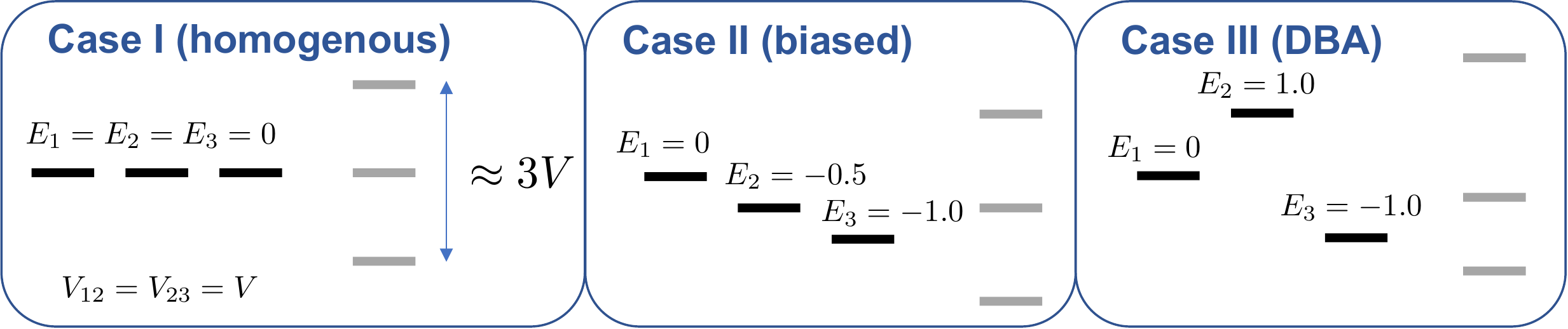}

\caption{Schematic representation of the trimer (three-level) systems evaluated in our study. Black bars indicate the local energies, whereas grey bars depict the (approximate) adiabatic energies. In the following, Case 1 refers to the homogeneous trimer, Case 2 refers to the biased trimer, and Case 3 refers to the donor--bridge--acceptor (DBA) trimer.}
\label{fig:trimers}
\end{figure}

We consider three different systems, schematic representation of which is shown in Figure \ref{fig:trimers}.
Case 1 is a homogeneous trimer characterized by the equal diagonal energies; Case 2 is biased trimer with the gradient diagonal energies; and Case 3 represents a donor-bridge-acceptor (DBA) trimer.

A systematic survey of the trimer system is presented below. 
All energies are given in the reference units of 208.5 cm$^{-1}$ (thermal energy at 300 K). The results from C-FSSH are compared against those from FSSH under Boltzmann sampling [FSSH (B)] and Wigner sampling [FSSH (W)] as well as numerically-exact results from the hierarchical equation of motion (HEOM). The surface hopping calculations were averaged over 2000 trajectories.
The convergence for HEOM results was insured in all system parameters. At 300 K, the hierarchy depth of 20 and a single Matsubara term were included. At 30 K, the hierarchy depth is set to 6 and 7 Matsubara terms were included.

\newpage

\begin{figure}[htbp]
\centering

\includegraphics[width=1.0\textwidth]{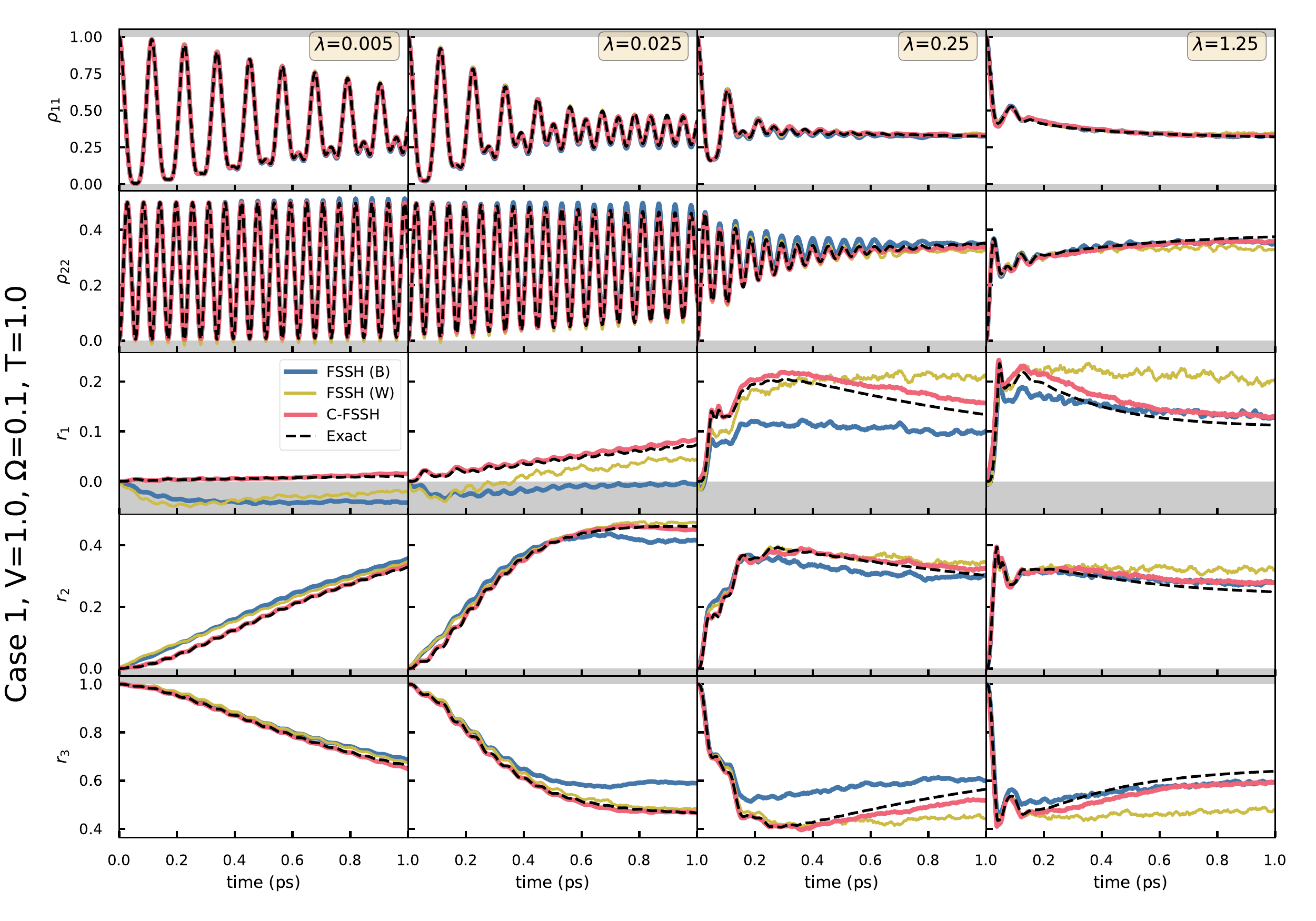}
\includegraphics[width=1.0\textwidth]{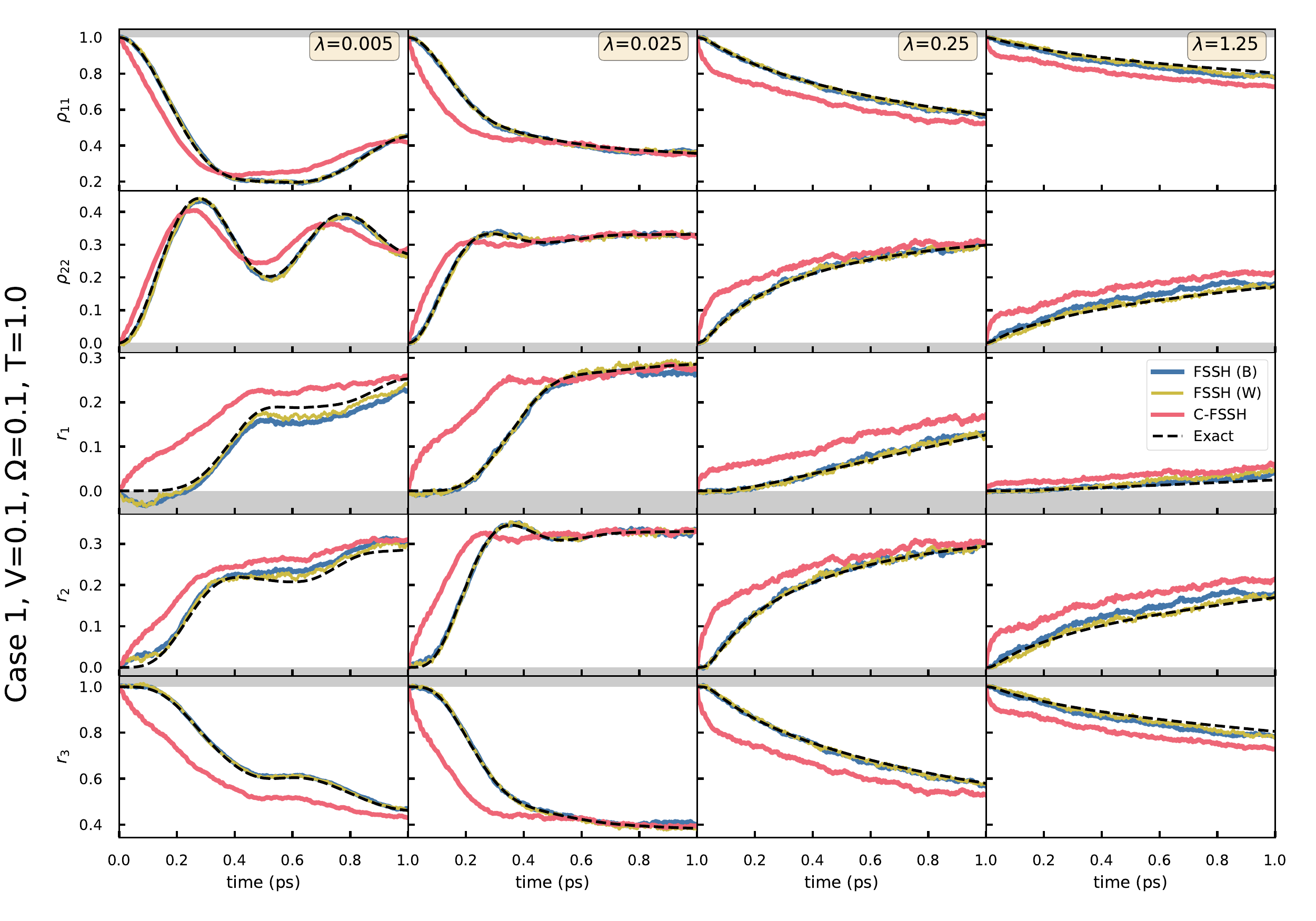}

\end{figure}

\newpage

\begin{figure}[htbp]
\centering

\includegraphics[width=1.0\textwidth]{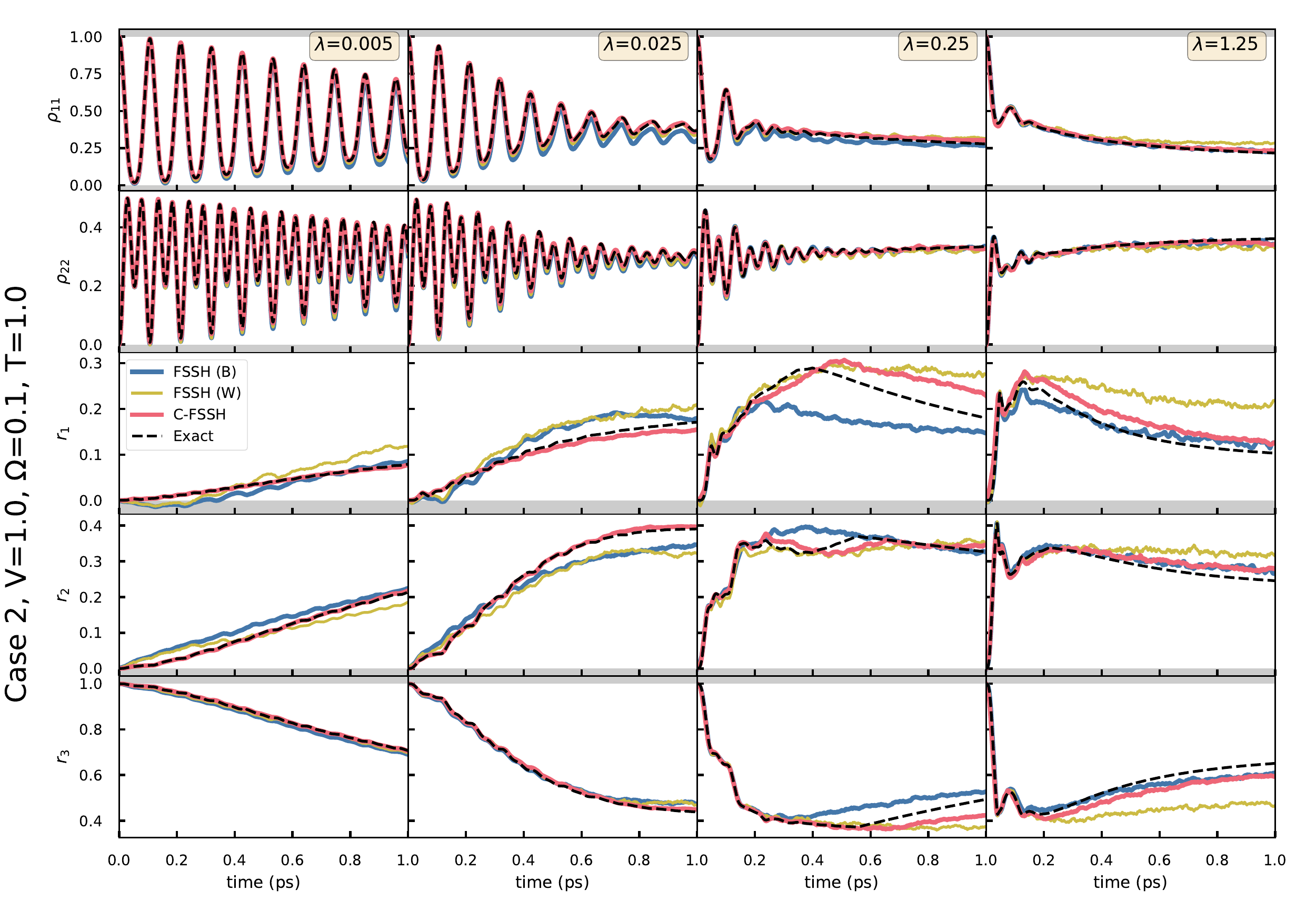}
\includegraphics[width=1.0\textwidth]{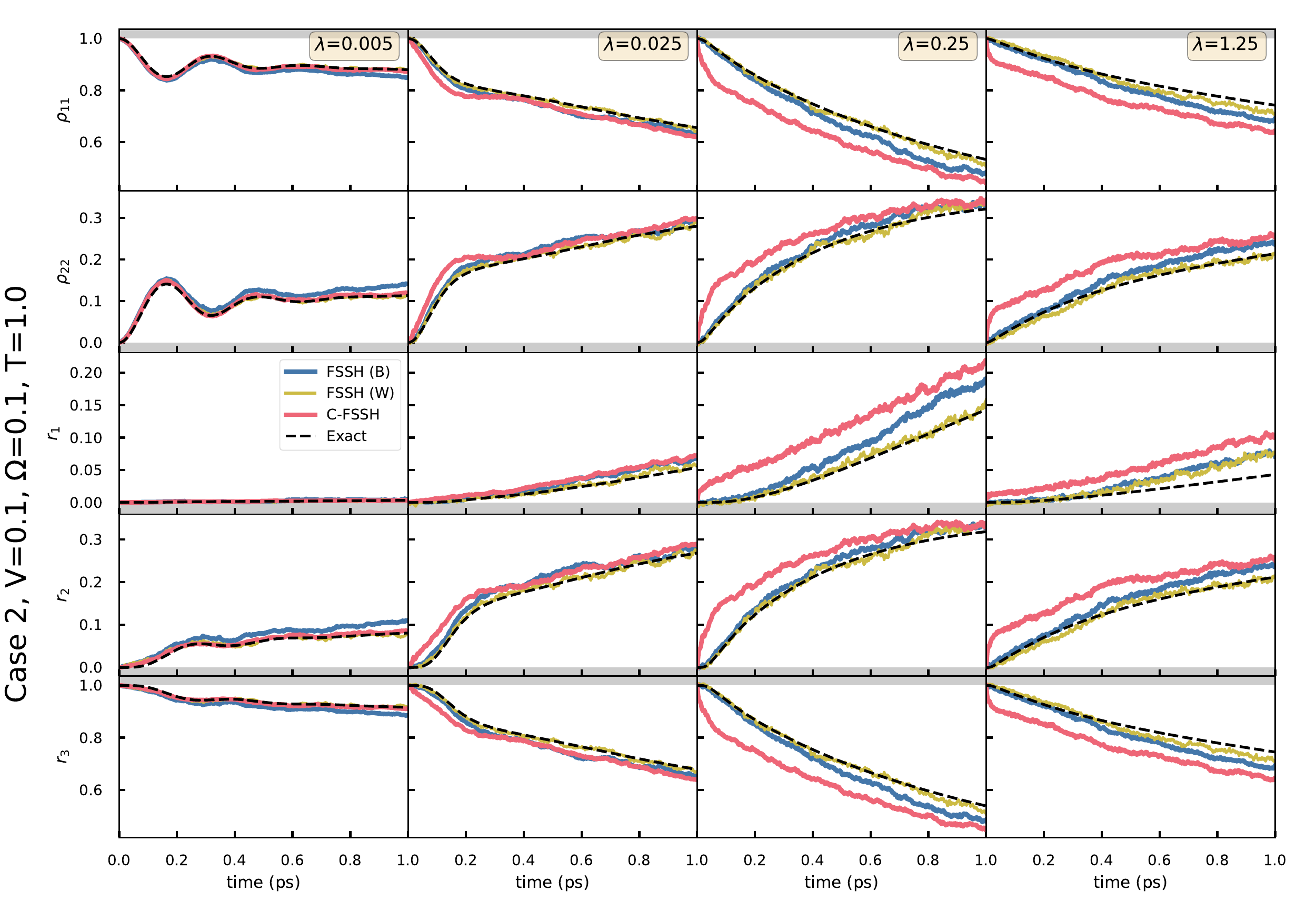}

\end{figure}

\newpage

\begin{figure}[htbp]
\centering

\includegraphics[width=1.0\textwidth]{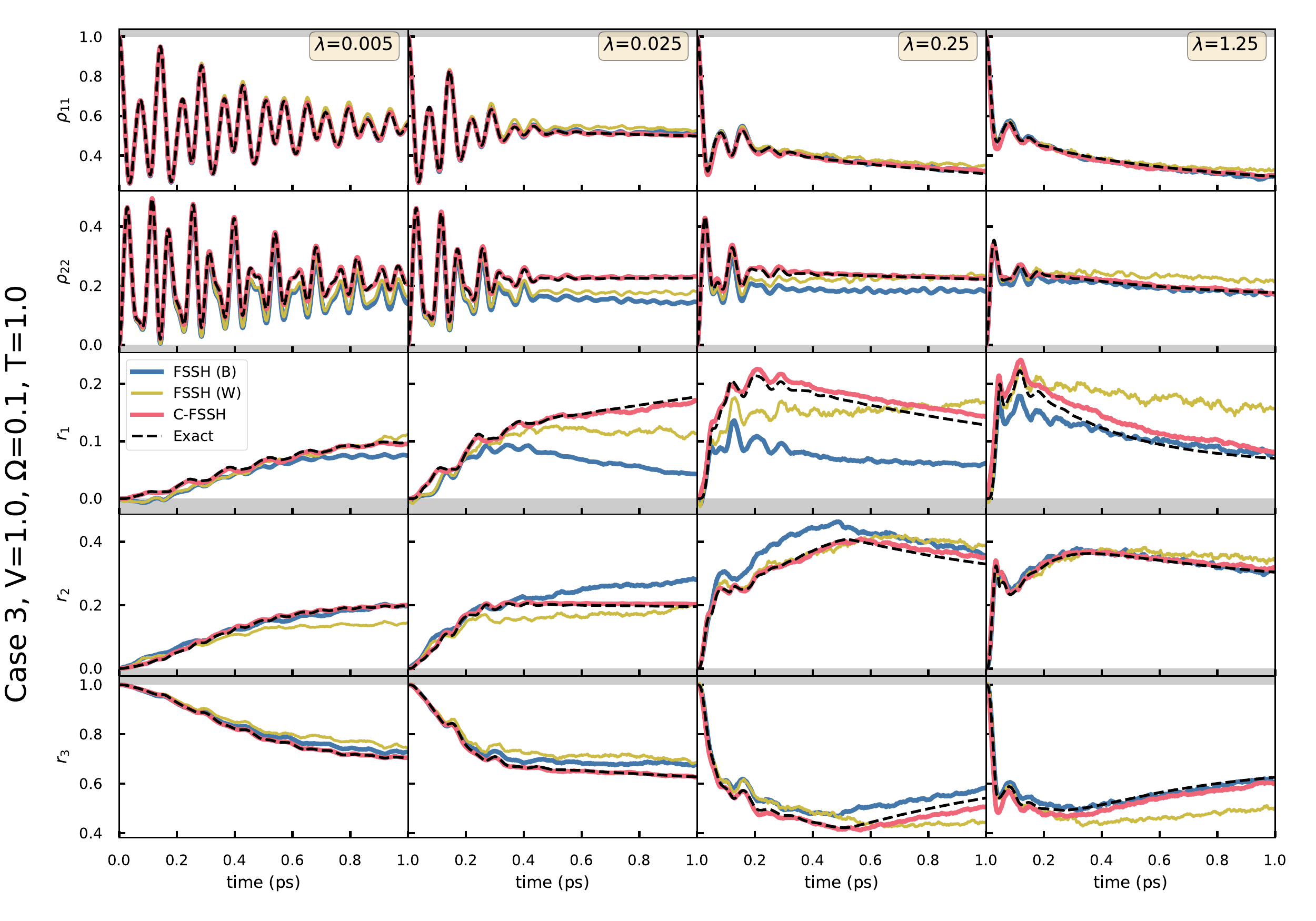}
\includegraphics[width=1.0\textwidth]{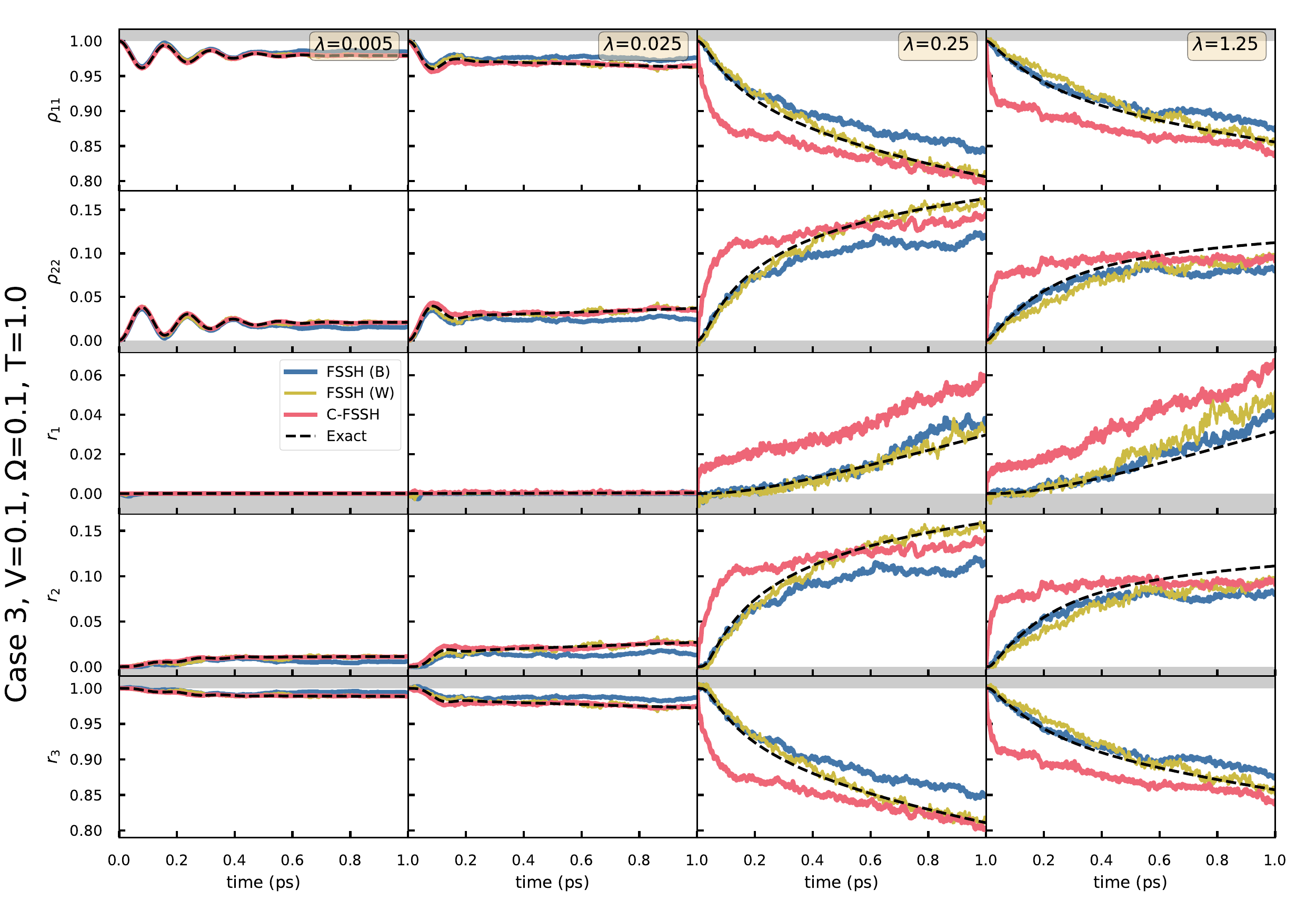}

\end{figure}


\begin{figure}[htbp]
\centering

\includegraphics[width=1.0\textwidth]{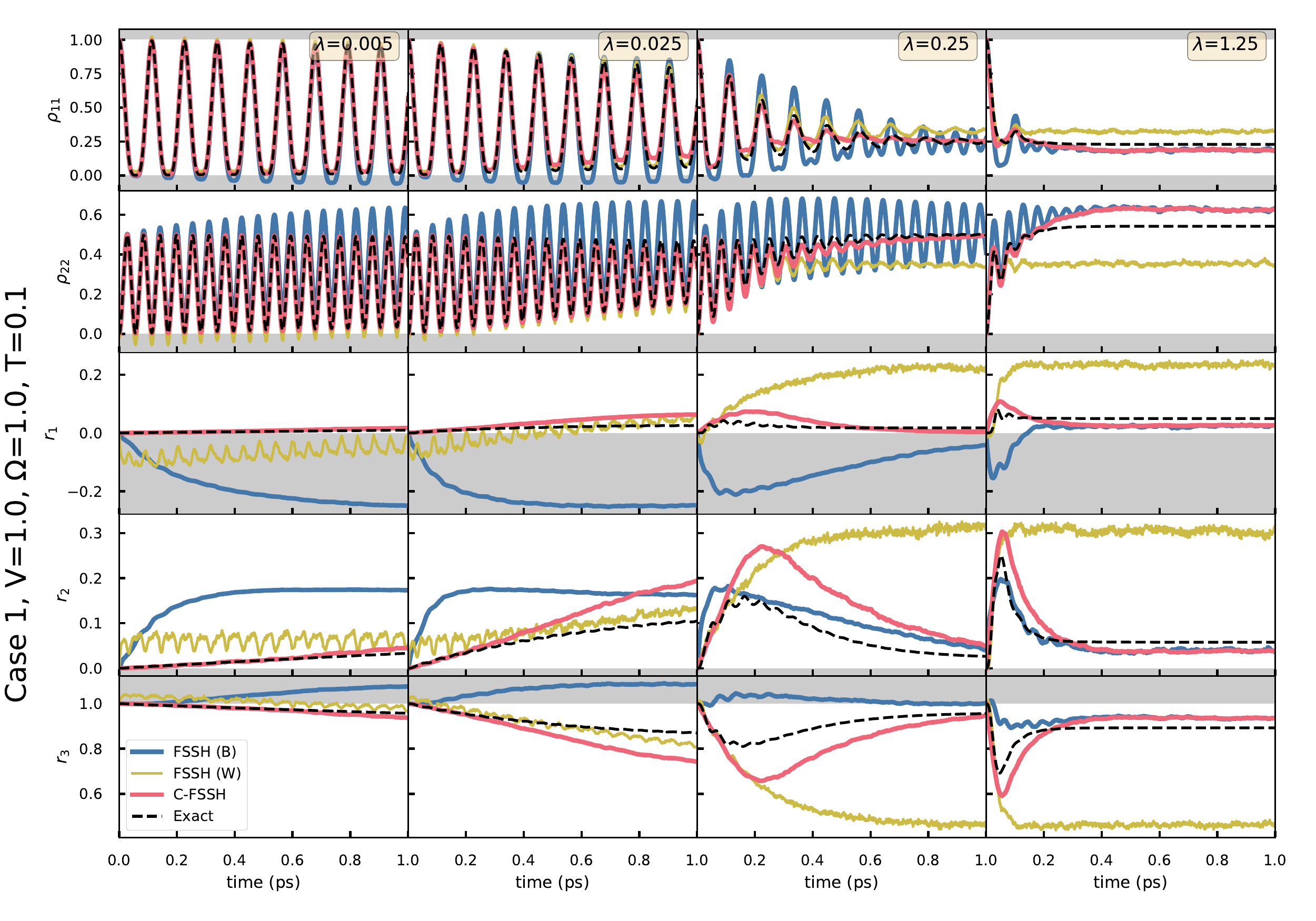}
\includegraphics[width=1.0\textwidth]{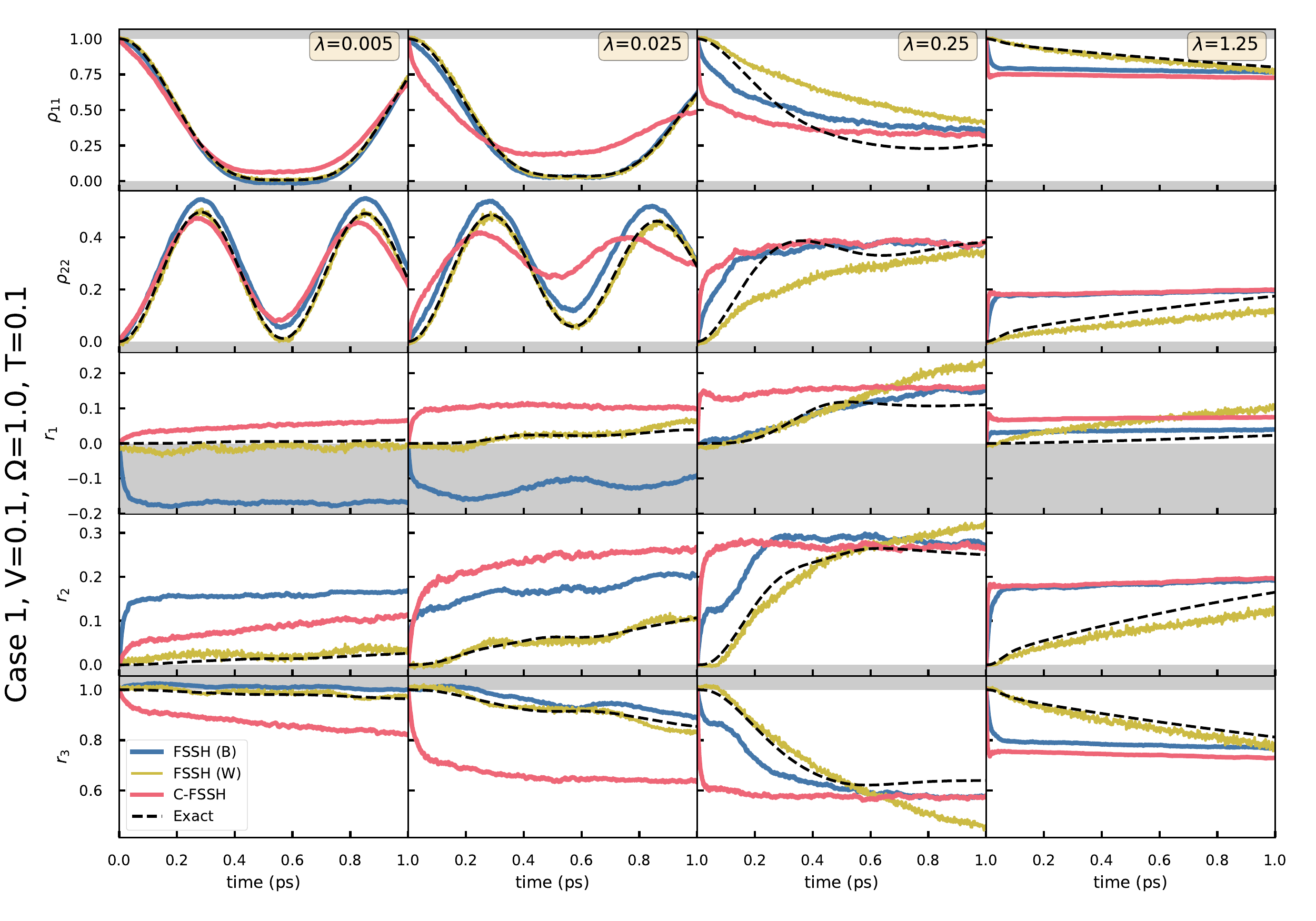}

\end{figure}

\newpage

\begin{figure}[htbp]
\centering

\includegraphics[width=1.0\textwidth]{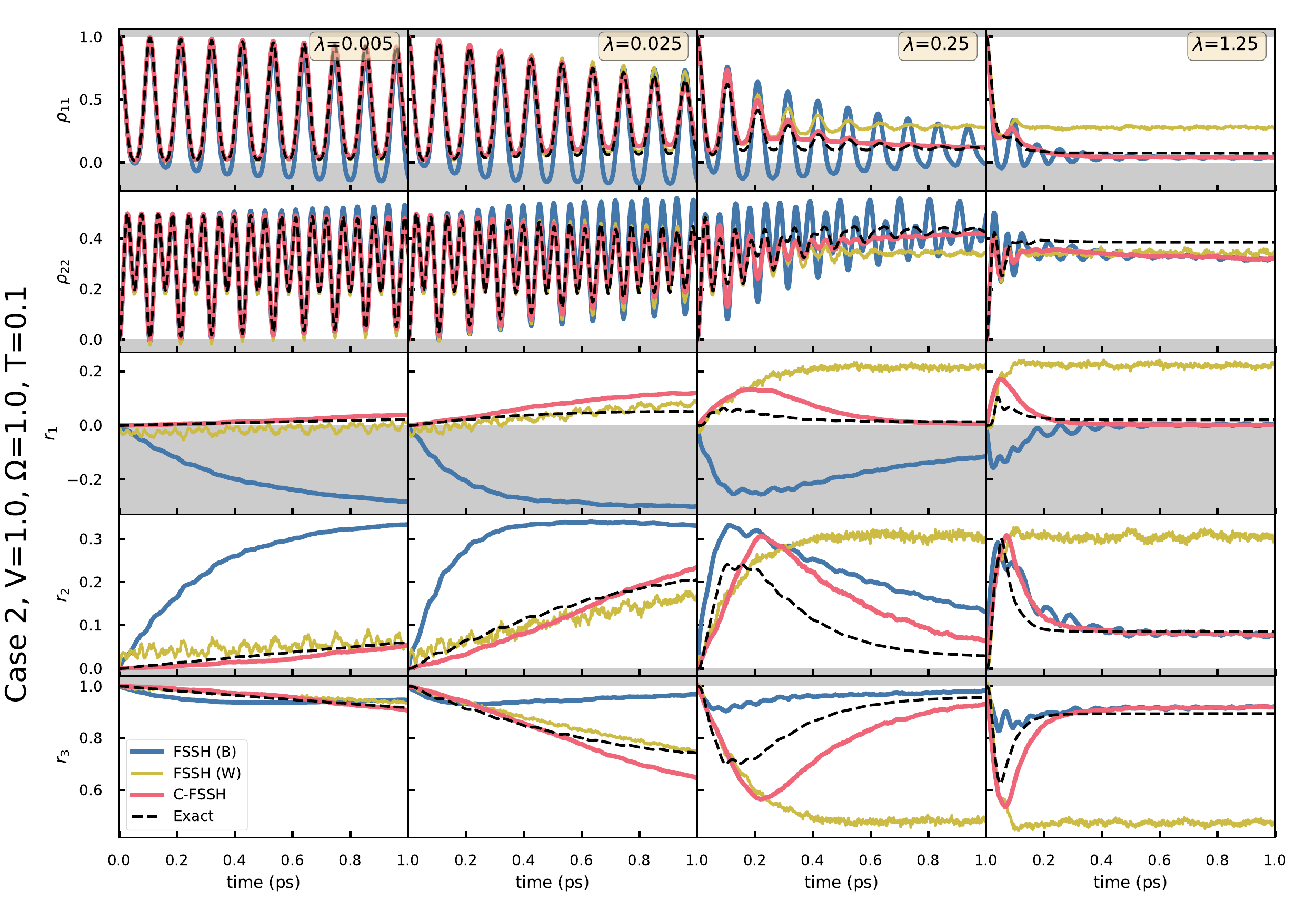}
\includegraphics[width=1.0\textwidth]{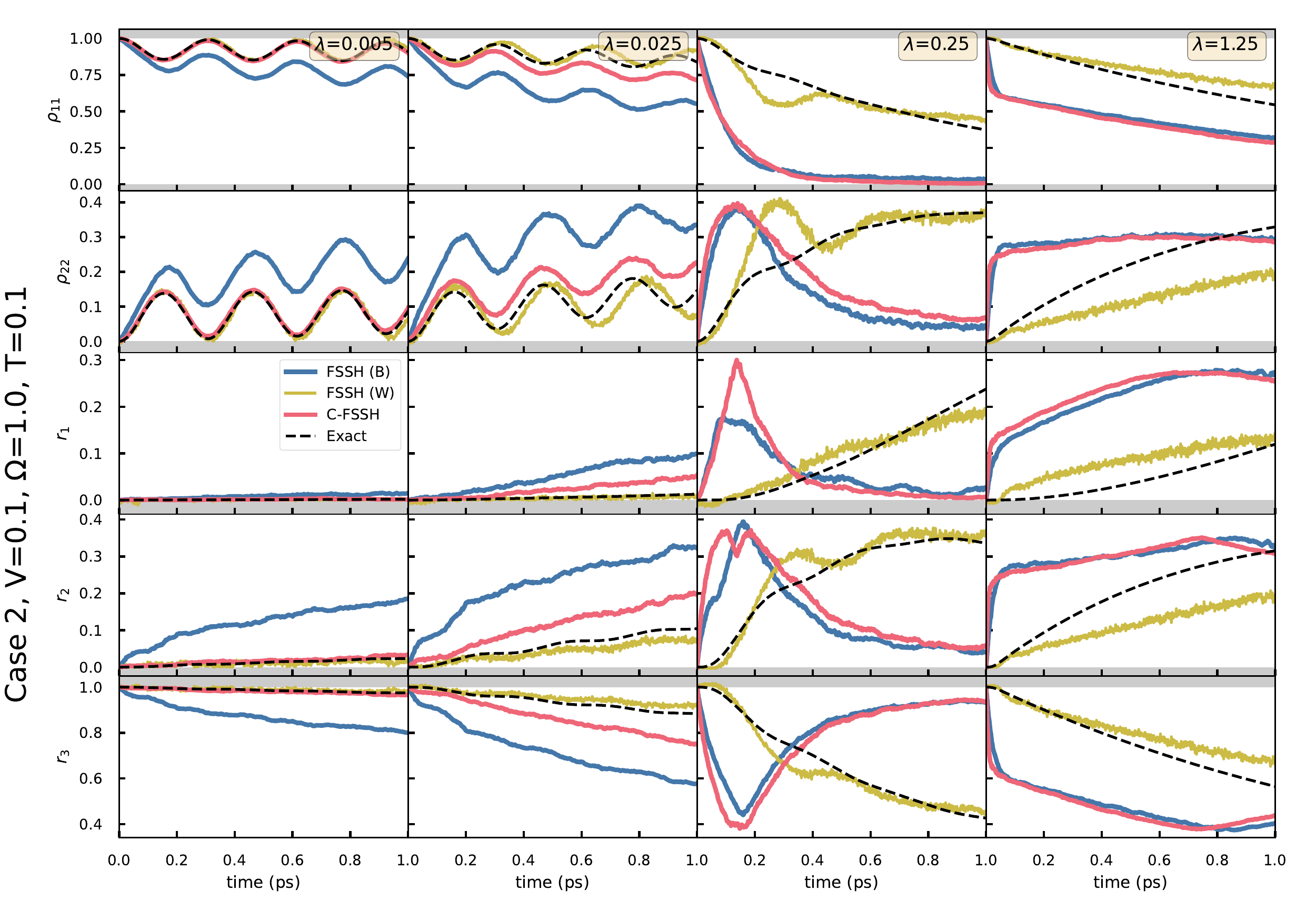}

\end{figure}

\newpage

\begin{figure}[htbp]
\centering

\includegraphics[width=1.0\textwidth]{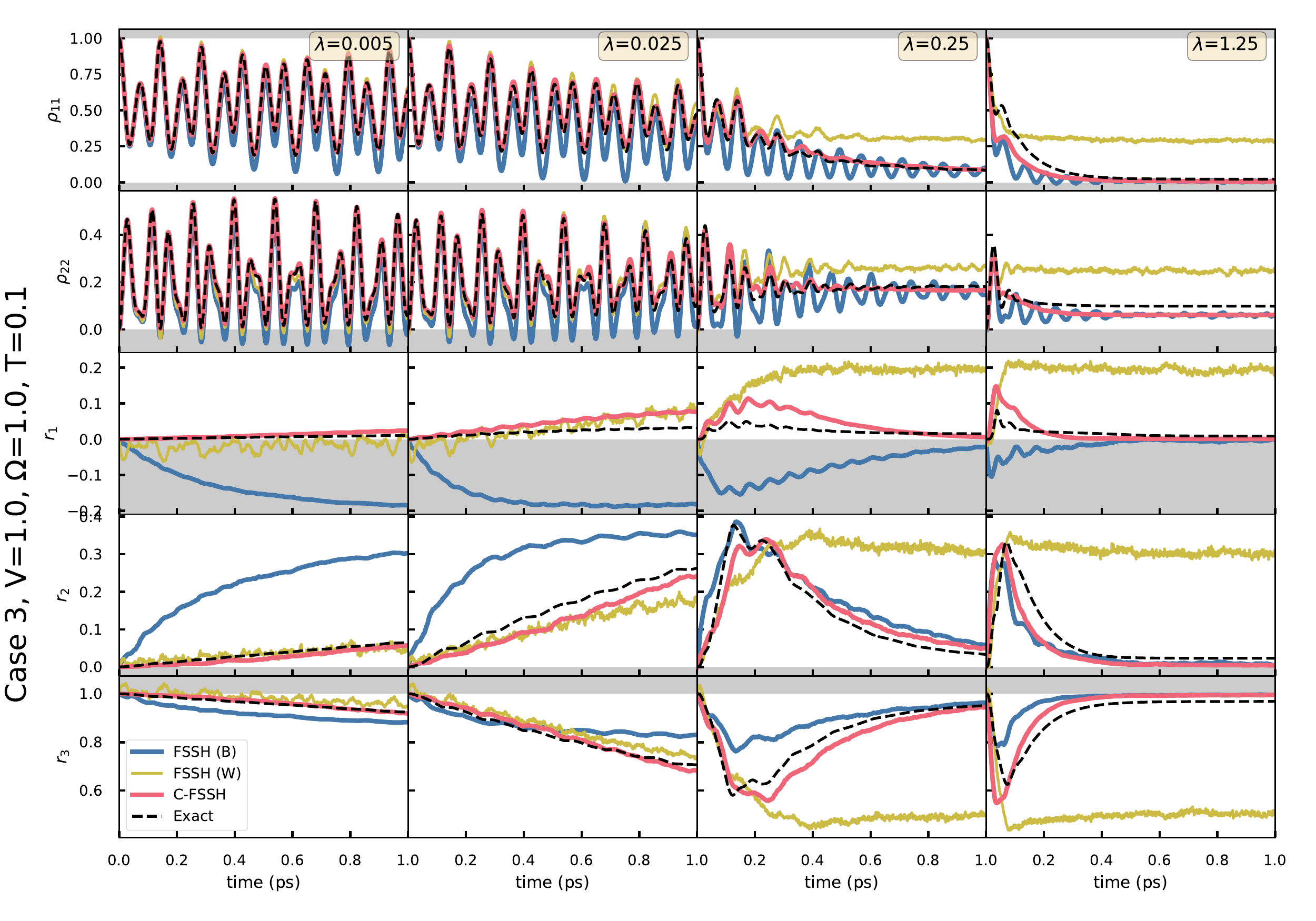}
\includegraphics[width=1.0\textwidth]{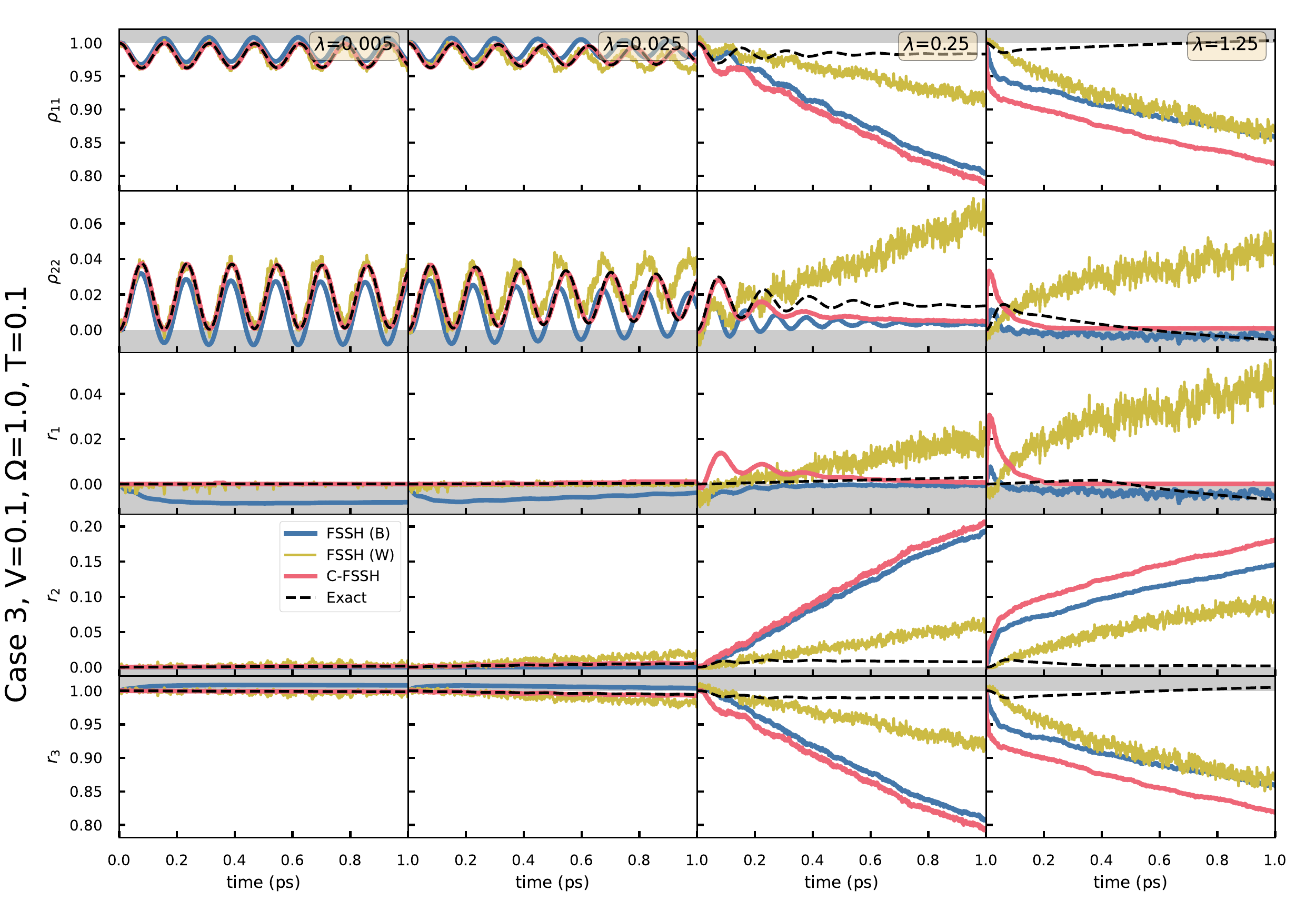}

\end{figure}
\bibliography{ankabib_extracted}